\documentclass[a4paper,11pt]{article}
\pdfoutput=1
\usepackage{jcappub} 
\usepackage[T1]{fontenc}
\usepackage{graphicx}
\usepackage{lipsum}
\usepackage{placeins}

\title{Testing signatures of phantom crossing through full-shape galaxy clustering analysis}

\author[a,1]{Emanuelly Silva,\note{Corresponding author.}}
\author[a,b]{Rafael C. Nunes}

\affiliation[a]{Instituto de Física, Universidade Federal do Rio Grande do Sul, 91501-970 Porto Alegre RS, Brazil}
\affiliation[b]{Divisão de Astrofísica, Instituto Nacional de Pesquisas Espaciais, Avenida dos Astronautas 1758, São José dos Campos, 12227-010, São Paulo, Brazil}

\emailAdd{emanuelly.santos@ufrgs.br}
\emailAdd{rafadcnunes@gmail.com}

\abstract{Recent observations of baryon acoustic oscillations (BAO) from the Dark Energy Spectroscopic Instrument (DESI) survey, when combined with measurements of the cosmic microwave background (CMB) and Type Ia supernovae (SNIa), provide compelling evidence for a phantom crossing at late times, along with statistically significant deviations from the standard $\Lambda$CDM model. In this work, we investigate the role of redshift-space galaxy clustering data by employing the pre-reconstruction full-shape (FS) galaxy power spectrum from the Baryon Oscillation Spectroscopic Survey (BOSS) data release 12 (DR12) sample. This dataset is analyzed in combination with BAO measurements from DESI data release 2 (DR2) and various SNIa samples. Our analysis demonstrates that the joint combination of these datasets can yield deviations from $\Lambda$CDM at a significance level of up to $\sim 5\sigma$, suggesting strong indications that the dark energy equation of state parameter $w(z)$ may have crossed the phantom divide ($w = -1$) in the redshift range $z \sim 0.4$–$0.5$. The precise location and strength of this crossing depend on the adopted theoretical parameterizations. Importantly, our results reveal that this trend persists even in the absence of CMB data, underscoring the robustness of the FS power spectrum as a powerful and independent probe for testing dark energy models and for distinguishing between competing cosmological scenarios.}

\begin{document}
\maketitle
\flushbottom

\section{Introduction}
\label{sec:intro}

The nature of dark energy (DE), responsible for the accelerated expansion of the universe, remains one of the central problems in modern cosmology. In the standard $\Lambda$CDM model, this phenomenon is attributed to a cosmological constant with an EoS parameter $w = -1$, corresponding to a uniform energy density in space and time. However, recent observations from the Dark Energy Spectroscopic Instrument (DESI) have placed this assumption under increasing pressure~\cite{DESI:2025zgx, DESI:2025zpo, DESI:2025fii, DESI:2024aqx}. By measuring baryon acoustic oscillations (BAO) in a record sample of over 14 million galaxies and quasars, DESI’s second data release reveals statistically significant deviations from $\Lambda$CDM predictions, with tensions reaching up to $4.2\sigma$, depending on the sample considered~\cite{DESI:2025zgx}. These results mark a turning point, suggesting that DE might not be a mere cosmological constant.

The growing number of tensions with $\Lambda$CDM—now also observed in BAO measurements—has intensified the need to explore complementary late-time probes capable of testing the standard model independently. In this endeavor, joint analyses combining the cosmic microwave background (CMB) with low-redshift observables, such as BAO or supernovae, have become a widely adopted strategy \cite{Silva:2025hxw, DESI:2025fii, Gialamas:2025pwv, Li:2025ula, Li:2025eqh, Lin:2025gne, Wolf:2025jed, Yashiki:2025loj, RoyChoudhury:2025dhe, Yang:2025ume, Giare:2024oil, Sabogal:2025mkp, Giare:2024smz, Wang:2025qpj, Liu:2025myr, Hogas:2025ahb, Zhou:2025kws, Pan:2025qwy, Escamilla:2025imi, Escamilla:2024xmz, Li:2025ops, Odintsov:2025jfq, Cheng:2025lod, Kumar:2025etf, Shah:2025ayl, Anchordoqui:2025fgz, Li:2024qso}. These combinations leverage the strengths of each probe, yielding stringent constraints on cosmological parameters across cosmic time. However, the CMB itself faces notable limitations on large angular scales ($\ell < 30$), where anomalies persist. These include (i) a low-power deficit in the temperature spectrum \cite{Planck:2018vyg, Billi:2023liq, Billi:2019vvg, Jung:2024slj}, and (ii) the lensing anomaly, where the smoothing of acoustic peaks exceeds theoretical expectations at the $2.8\sigma$ level \cite{Domenech:2020qay, Addison:2023fqc, Renzi:2017cbg, DiValentino:2019qzk, DiValentino:2022oon, DiValentino:2020hov}. While such features may be statistical in nature \cite{Planck:2019evm, Planck:2019nip}, their unresolved origin highlights the importance of obtaining independent constraints free from CMB assumptions.

In this context, the full-shape (FS) galaxy power spectrum stands out as a powerful and relatively underexplored probe. Unlike geometric observables, which constrain the background expansion, FS measurements capture the scale-dependent clustering of matter and are directly sensitive to the physics of structure growth. Recent advances, particularly within the framework of the Effective Field Theory of Large-Scale Structure (EFTofLSS) \cite{Baumann:2010tm, DAmico:2019fhj, Carrasco:2012cv, Carrasco:2013mua, Simonovic:2017mhp}, enable accurate modeling of this observable by treating galaxies as a biased tracer of a nonlinear, imperfect fluid. The effective stress tensor introduced by this theory encapsulates the influence of short-scale dynamics via counterterms, which preserve symmetries and are calibrated from data. This approach allows for robust control over theoretical uncertainties and significantly extends the reach of perturbation theory into mildly nonlinear regimes \cite{Chudaykin:2020aoj}. As a result, FS analyses can simultaneously constrain $\sigma_8$, the growth rate via redshift-space distortions (RSD), and the shape of the primordial power spectrum, while enhancing sensitivity to scale-dependent effects predicted by extensions of $\Lambda$CDM, such as modified gravity or dynamical DE \cite{Chen:2024vuf, Shim:2024pce, Noriega:2024lzo, Holm:2023laa, Ivanov:2023qzb, Simon:2022hpr, Simon:2022csv, Simon:2022lde, Simon:2022adh, Noriega:2022nhf, Carrilho:2022mon, Tsedrik:2022cri, Tanseri:2022zfe, Kumar:2022vee, DAmico:2019fhj, Colas:2019ret, Zhang:2021yna, DAmico:2020kxu, Chudaykin:2024wlw, Noriega:2024eyu, Philcox:2022sgj, Rodriguez-Meza:2023rga, Semenaite:2022unt}. Recent studies have discussed FS analyses based on mock data in the context of DESI samples \cite{DESI:2024jis, DESI:2024hhd, DESI:2025wzd, NovellMasot:2025fju}.

These robust analyses and constraints are particularly pertinent in the current era of precision cosmology, characterized by big volumes of high-quality data and steadily diminishing systematic uncertainties across a variety of observational probes. This improved accuracy has led not only to tighter parameter constraints within the $\Lambda$CDM framework but also to the emergence of statistically significant tensions between different datasets, potentially signaling the need for new physics beyond the standard model. The most prominent of these is the Hubble tension: a discrepancy exceeding $5\sigma$~\cite{Riess:2021jrx, DiValentino:2024yew, Perivolaropoulos:2024yxv} between the value of the Hubble constant $H_0$ inferred from early-universe measurements (e.g., Planck CMB data~\cite{Planck:2018vyg}) and that obtained from local distance ladder observations by SH0ES~\cite{Riess:2021jrx}. A second key discrepancy involves the parameter $S_8 \equiv \sigma_8 \sqrt{\Omega_{\rm m}/0.3}$, which quantifies the amplitude of matter density fluctuations on $8\,h^{-1}\,\mathrm{Mpc}$ scales. While recent weak lensing results, notably from KiDS-1000~\cite{Wright:2025xka}, now show good agreement with Planck-CMB predictions, FS galaxy clustering analyses continue to report a significant tension, up to $4.5\sigma$~\cite{Ivanov:2024xgb, Chen:2024vuf}, relative to early-universe expectations. Additional hints of discrepancy in $S_8$ also arise from RSD measurements~\cite{Nunes_Vagnozzi_2021, Kazantzidis_2018} and other late-time probes~\cite{Karim:2024luk, Dalal:2023olq}. Taken together, these tensions—originating from independent and complementary data sets—have motivated extensive investigations into extensions of the standard model, including evolving DE models \cite{Shajib:2025tpd}, interactions with dark matter \cite{Silva:2024ift}, and various modifications of gravity \cite{Chudaykin:2025gdn} (see~\cite{CosmoVerse:2025txj} for a review).

In this work, we aim to provide robust constraints on two distinct dynamical DE parametrizations, independently of CMB data. To achieve this, we combine a FS analysis of the galaxy power spectrum using data from the Baryon Oscillation Spectroscopic Survey Data Release 12 (BOSS DR12) with the most up-to-date supernova samples and the latest geometric BAO measurements from the second data release of DESI (DESI-DR2). Our analysis is centered on investigating the dynamical behavior of DE at late times, with particular attention to recent discussions surrounding a possible phantom crossing. Importantly, our approach is largely independent of CMB data, emphasizing instead the constraining power of low-redshift observables. In addition to probing the nature of DE, we examine the implications of these models for current cosmological tensions, specifically those related to the $H_0$ and $S_8$. \textit{As we will demonstrate, the FS analysis, when combined with geometric distance measurements, constitutes a robust and sensitive cosmological probe. Notably, our results provide strong statistical support for a phantom crossing occurring at redshift $z \sim 0.5$, offering insights into potential deviations from the standard $\Lambda$CDM model.}

The \textit{paper} is organized as follows. In Section~\ref{sec:model}, we introduce the two DE parametrizations considered and briefly discuss their impact on the galaxy power spectrum, with emphasis on the theoretical framework provided by the EFTofLSS. Section~\ref{sec:methodology} details the analysis pipeline, the observational datasets employed, and the statistical methods used. In Section~\ref{sec:results}, we present and discuss our main results, comparing the performance of both parametrizations across the various data combinations. Finally, Section~\ref{sec:finalremarks} summarizes our findings and outlines directions for future investigation.

\section{Dynamical Dark Energy and the full-shape galaxy power spectrum}
\label{sec:model}

In this section, we present the dynamical DE parameterizations explored in this work, together with an overview of the EFTofLSS, which provides the foundation for modern FS analyses. We also discuss how the evolution of cosmic structures is affected when the assumption of a constant DE equation-of-state (EoS) is relaxed in favor of time-dependent models.

\subsection{The $w_0$–$w_a$ parametrization}
The Chevallier–Polarski–Linder (CPL) model, proposed independently by \cite{Chevallier2001} and \cite{Linder2003}, will be the first DE parameterization examined in this work. It was introduced as a phenomenological framework to account for possible time dependencies in the DE EoS. The model remains widely used in the literature due to its parametric simplicity—requiring only two degrees of freedom ($w_0$, $w_a$)—and its phenomenological robustness, being capable of describing smooth temporal variations of $w(z)$ without introducing physical singularities at any redshift.

The CPL parametrization expresses $w(z)$ as a first-order Taylor expansion around the present epoch. In terms of the scale factor, $a = 1/(1 + z)$, it takes the form
\begin{equation}
    w(a) = w_0 + w_a (1 - a),
\end{equation}
where $w_0$ denotes the present-day value of the EoS, and $w_a$ quantifies its first-order variation with respect to the scale factor. Physically, $w_a$ governs the dynamical evolution of DE during cosmic expansion. If $w_a > 0$, the DE density decreases with time. Conversely, if $w_a < 0$, the DE density grows over time, suggesting that DE becomes increasingly dominant in the energy budget of the universe.

This expression guarantees regular behavior throughout the entire cosmic history, avoiding unphysical divergences both in the early universe ($z \to \infty$, where $w \to w_0 + w_a$) and in the far future ($z \to -1$, where $w \to w_0$). An important feature of the CPL parameterization is its ability to recover the standard $\Lambda$CDM model as a limiting case when $w_0 = -1$ and $w_a = 0$. This allows for direct and systematic tests of the cosmological constant hypothesis within a broader dynamical DE framework.

In general terms, several other two-parameter DE parameterizations involving $w_0$ and $w_a$ have been proposed in the literature \cite{Barboza:2008rh, Dimakis:2016mip, Jassal:2005qc, Efstathiou:1999tm}. However, from the perspective of observational constraints, these models are practically equivalent in their predictions \cite{Giare:2024gpk}. Therefore, without loss of generality, employing the CPL parametrization is more than sufficient for the purposes of this work, as it captures the essential phenomenology while avoiding unnecessary parametric redundancies.

\begin{figure}[htpb!]
    \centering
    \includegraphics[width=\columnwidth]{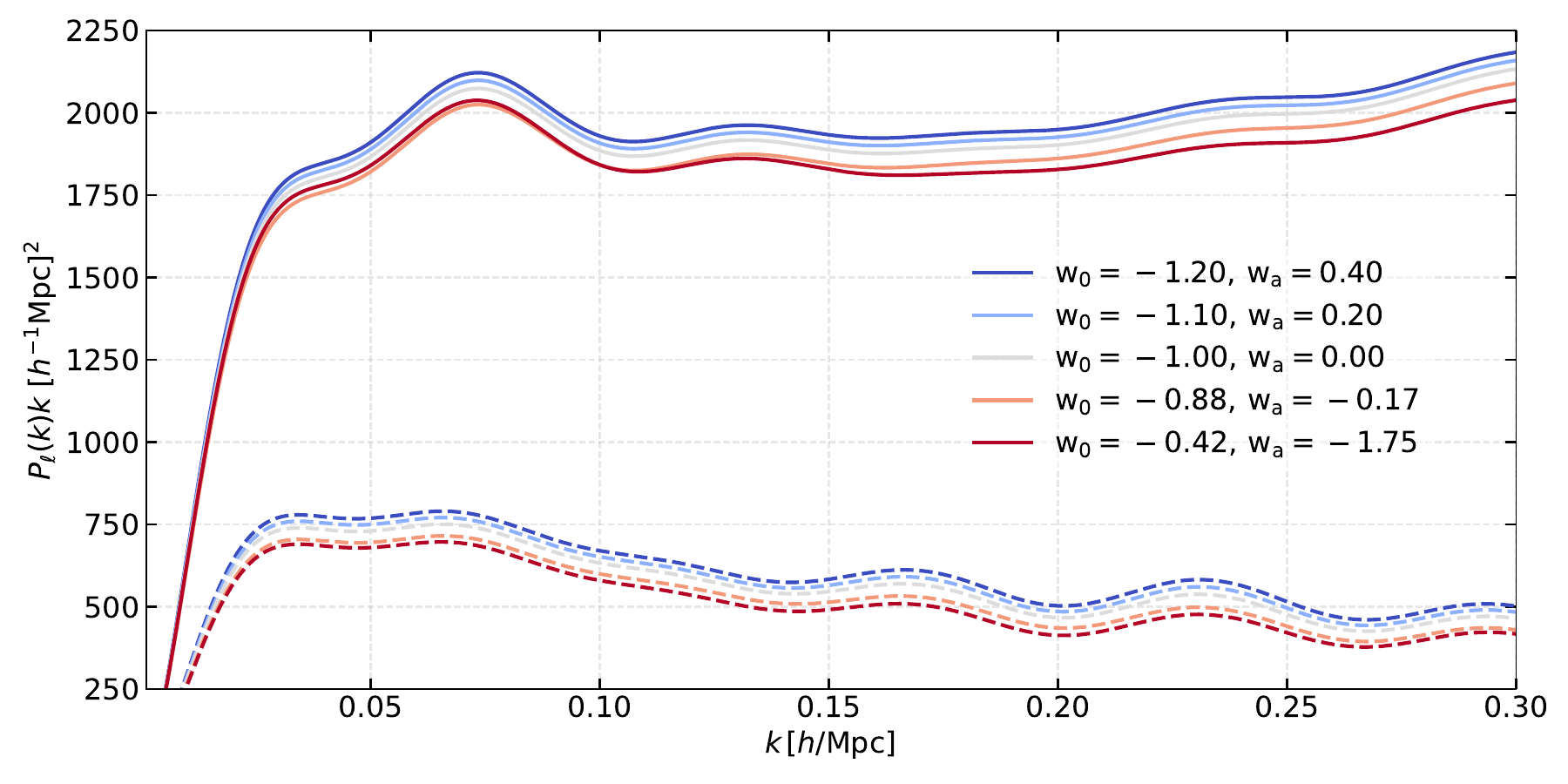} 
    \caption{Galaxy power spectrum multipoles: monopole ($\ell = 0$, solid lines) and quadrupole ($\ell = 2$, dashed lines). We vary the values of the DE parameters $w_0$ and $w_a$, as indicated in the legend, showing that different choices lead to significant changes in the galaxy power spectrum.}
    \label{fig:wa_w0_vary}
\end{figure}

In Figure~\ref{fig:wa_w0_vary}, we illustrate the quantitative impact of varying the DE EoS parameters $w_0$ and $w_a$ on the multipoles of the galaxy power spectrum. All other cosmological parameters are held fixed at their Planck-$\Lambda$CDM best-fit values \cite{Planck:2018vyg}. The results show that the choice of $w_0$ and $w_a$—which determines whether the DE behavior resembles a quintessence-like or phantom scenario—can significantly affect the amplitude of the power spectrum multipoles $P_\ell(k)$, particularly the monopole term. In addition to changes in amplitude, subtle modifications to the overall shape of the spectra are also observed, especially on large scales (low $k$), although amplitude variations remain the dominant effect. 

\subsection{The $w_{\dagger}$CDM parametrization}

A recent work by \cite{Scherer:2025esj} proposed a particularly interesting test which, in addition to constraining the EoS, investigates the possibility of a transition between quintessence and phantom regimes—or vice versa, drive only by data only preference. On the other hand, this simple dynamical test can be theoretically mapped in a robust manner onto the framework VCDM model \cite{DeFelice:2020eju}, which represents a class of models capable of accommodating a wide range of time-dependent DE behaviors. Importantly, these models do not introduce any additional propagating degrees of freedom in the gravitational or matter sectors and remain free from background-level theoretical inconsistencies.
 
In general terms, it is proposed that the EoS undergoes a transition at a certain critical redshift, characterized by a parameter $\Delta$ that controls the amplitude of the transition. This change is centered around a critical scale factor $a_{\dagger}$ (corresponding to redshift $z_{\dagger}$), and models a sudden shift in the value of $w$. The corresponding phenomenological model is given by:
\begin{equation}
\label{w_new}
w(a) = -1 + \Delta \, \operatorname{sgn}(a_{\dagger} - a) \, ,
\end{equation}
where $\operatorname{sgn}$ denotes the sign function. This form captures a transition in the DE dynamics, switching from $w = -1 + \Delta$ at early times ($a < a_{\dagger}$) to $w = -1 - \Delta$ at late times ($a > a_{\dagger}$), or vice versa, depending on the sign of $\Delta$. The corresponding smooth functional form of the EoS is provided in equation (8) of \cite{Scherer:2025esj}.

This parametrization introduces a discontinuous behavior for $w(a)$, which is not constrained by any fixed statistical prior, but is instead fully determined by the data through the statistical evolution of the Markov chains during the fitting procedure (see \cite{Scherer:2025esj}). It is important to emphasize that this parameterization is specifically designed to capture potential transitions between quintessential and phantom regimes. In other words, the framework does not assume a fixed dynamical class for dark energy but instead encodes the possibility that the equation of state may cross the cosmological constant boundary during cosmic evolution. The model itself does not prescribe when or how this transition occurs; rather, both the timing and the detailed nature of the crossing are determined entirely by observational constraints. By construction, this feature makes the parameterization fundamentally incompatible with thawing-type scenarios, where the dynamics are tied to a monotonic evolution of the scalar field. For dedicated analyses of thawing behaviors, see \cite{Wolf:2024eph, Dinda:2025iaq, Bayat:2025xfr}. As demonstrated in \cite{Scherer:2025esj}, and further reinforced by the results presented below, the statistical preference for our proposal is significantly stronger than for thawing-type dynamics, underscoring the relevance of a phantom-crossing description. We emphasize that the $w_{\dagger}$CDM parametrization is purposefully designed to capture relevant trends, particularly the possible crossing of the phantom divide, thereby providing tighter constraints on the parameters.

\subsection{Effective Field Theory of Large-Scale Structure for DDE}
\label{EFT}

A FS analysis of the galaxy power spectrum provides access to cosmological information beyond the BAO features, but fully exploiting the richness of this observable requires a theoretical framework capable of reliably extending into progressively more nonlinear regimes — effectively capturing a larger number of Fourier modes. Historically, standard perturbation theory (SPT) \cite{Scoccimarro:1995if, Bernardeau:2001qr} has been the primary tool for addressing nonlinear scales. However, higher-order corrections within SPT failed to yield substantial improvements in predictive power, motivating alternative approaches. In this context, the EFTofLSS was developed, recognizing that the perfect fluid approximation becomes inherently inconsistent beyond very large scales. Instead, EFTofLSS treats each component of the cosmic energy density (e.g., dark matter and baryons) as a separate field, subject to corrections arising from imperfect fluid dynamics. In this subsection, we provide a brief theoretical overview of the EFTofLSS framework adopted in this work and highlight its implications for DDE models. For a comprehensive exposition of the formalism and its foundational aspects, we refer the reader to \cite{Carrasco:2012cv, Carrasco:2013mua, Baumann:2010tm, Bernardeau:2001qr, Ivanov:2022mrd, Crocce:2005xy, Bernardeau:2008fa, Bernardeau:2011dp, Hand:2017ilm, Cabass:2022avo, Green:2024cmx, Assassi:2015jqa, Baldauf:2020bsd, Blas:2013bpa, Senatore:2017hyk, Ivanov:2018gjr, Lewandowski:2015ziq, Lewandowski:2014rca}.

The starting point of EFTofLSS is the Boltzmann equation, which, when solved with a focus on the dynamics of long-wavelength (small $k$) modes, naturally gives rise to imperfect fluid terms organized into a perturbative derivative expansion. The form of these terms is dictated by symmetry considerations, while their amplitudes remain free parameters that must be calibrated against data. These free parameters effectively encapsulate the impact of unresolved small-scale physics, substantially enhancing the performance of cosmological parameter constraints derived from large-scale structure observations \cite{Chudaykin:2020aoj}.

In the context of modeling the galaxy power spectrum—the primary observable investigated in this work—we employ the EFTofLSS framework to incorporate one-loop corrections. First, we include one-loop corrections to the matter power spectrum, which consist of SPT terms alongside an effective field theory counterterm that captures the influence of small-scale physics beyond perturbation theory. This counterterm introduces a scale-dependent correction proportional to $k^2 P_{\text{lin}}(k)$, whose amplitude is treated as a free parameter to be constrained by data.

Next, we include Alcock-Paczynski (AP) distortions by mapping between the true cosmology and the fiducial cosmology adopted to analyze the data. This mapping modifies both the observed wavenumbers $k$ and the cosine of the angle $\mu$ between the wavevector and the line-of-sight, thereby altering the shape and anisotropy of the measured galaxy power spectrum. The transformations are given by
\begin{align}
k^2_\mathrm{true} &= k^2_\mathrm{obs}\left[\left(\frac{H_\mathrm{true}}{H_\mathrm{fid}}\right)^2 \mu_\mathrm{obs}^2 
+ \left(\frac{D_{A,\mathrm{fid}}}{D_{A,\mathrm{true}}}\right)^2 (1 - \mu_\mathrm{obs}^2)\right],\\
\mu^2_\mathrm{true} &= \frac{\left(\frac{H_\mathrm{true}}{H_\mathrm{fid}}\right)^2 \mu_\mathrm{obs}^2}
{\left(\frac{H_\mathrm{true}}{H_\mathrm{fid}}\right)^2 \mu_\mathrm{obs}^2 
+ \left(\frac{D_{A,\mathrm{fid}}}{D_{A,\mathrm{true}}}\right)^2 (1 - \mu_\mathrm{obs}^2)},
\end{align}
where $H$ denotes the Hubble parameter and $D_A$ the angular diameter distance, with subscripts ``true'' and ``fid'' referring to the true and fiducial cosmologies, respectively.

For biased tracers, we employ a comprehensive bias expansion that captures the nonlinearities of galaxy formation through the parameters $b_1$, $b_2$, and $b_{\mathcal{G}_2}$. Specifically, we model the galaxy overdensity as
\begin{equation}
\delta_g(\vec{x}) = b_1 \delta(\vec{x}) + \frac{b_2}{2} \delta^2(\vec{x}) + b_{\mathcal{G}_2} G_2(\vec{x}),
\end{equation}
where $\delta$ is the matter overdensity and $G_2$ represents the tidal field contribution. In redshift space, we fully account for the angular dependence of RSD by incorporating the Kaiser effect along with velocity counterterms that describe small-scale velocity dispersions, often referred to as the fingers-of-God. To robustly capture the nonlinear damping of the BAO signal, we implement infrared (IR) resummation. This procedure separates the linear power spectrum into smooth and oscillatory (wiggly) components and applies a direction-dependent damping factor to the oscillatory part, effectively modeling the impact of large-scale bulk flows and mode coupling that broaden the BAO feature.

With respect to the kernels, it is important to note that their structure arises solely from the underlying gravitational dynamics and the equivalence principle \cite{Chudaykin:2020aoj}. Thus, while modifying the DE EoS affects the background expansion, the linear growth of structures, and the linear power spectrum, it does not change the fundamental way in which different Fourier modes interact nonlinearly. In other words, the mode-coupling kernels, which encode these nonlinear interactions, remain unchanged. Consequently, no modifications to the kernels are required in our analysis. This behavior stands in contrast to modified gravity models or scenarios that violate the equivalence principle, where new coupling terms arise in the kernels due to nonlinear Poisson corrections or scalar–tensor interactions.

For further technical details on all these corrections, we refer the reader to \cite{Chudaykin:2020aoj} and the references therein. Taking all the effects discussed above into account, we restrict our analysis to the monopole and quadrupole moments ($\ell=0,2$) of $P_{g,\ell}(k)$. These multipoles are computed from the anisotropic two-dimensional galaxy power spectrum via
\begin{equation}
P_{g,\ell}(k) = \frac{2\ell+1}{2} \int_{-1}^{1} d\mu\, P_g(k,\mu) P_{\ell}(\mu),
\end{equation}
where $\mu$ denotes the cosine of the angle between the Fourier mode $k$ and the line-of-sight direction, and $P_{\ell}(\mu)$ are Legendre polynomials of order $\ell$.

In Figure~\ref{fig:wa_w0_vary}, we present the galaxy power spectrum multipoles $P_{g,\ell}(k)$ considered in our analysis, where the monopole ($\ell=0$) is shown as solid lines and the quadrupole ($\ell=2$) as dashed lines. These theoretical outputs were produced using the \texttt{CLASS-PT} \cite{Chudaykin:2020aoj}. Examining the dynamical DE scenario under the CPL parametrization—which we systematically investigate in this work, it becomes evident that, as previously noted, variations in the DE parameters $w_0$ and $w_a$ do not significantly alter the overall shape of the galaxy power spectrum. The characteristic scale dependence associated with features such as the BAO remains largely unaffected across different values of these parameters. However, changes in $w_0$ and $w_a$ lead to pronounced variations in the amplitude of both multipoles. Physically, this is a direct consequence of how DE influences the growth history of cosmic structures and the overall normalization of matter fluctuations, encapsulated by parameters like $\sigma_8$. This behavior underscores the critical importance of analyses that exploit the full shape of the power spectrum, rather than relying solely on geometric BAO information. These variations highlight the sensitivity of $P_\ell(k)$ to the underlying DE dynamics.

In this work, we follow a phenomenological approach, adopting simple redshift-dependent parameterizations, $w(z)$, to model the dynamical behavior of DE, including the two cases considered within the scope of this study. The evolution of $w(z)$ is incorporated directly into the background and linear perturbation equations via a modified version of the Einstein-Boltzmann solver, ensuring full consistency between background dynamics and structure formation. This modification propagates through all computations tied to the expansion history and linear growth, affecting quantities such as the Hubble parameter and the linear matter power spectrum. Importantly, no changes to the internal routines that compute the perturbation theory kernels or response matrices within \texttt{CLASS-PT}~\cite{Chudaykin:2020aoj} were required, as these remain dictated solely by gravitational dynamics and the equivalence principle, as already justified above. Finally, we note that similar strategies—where new physics enters primarily through modifications to the linear power spectrum—have been widely employed in recent FS analyses of galaxy clustering data to constrain both early and late-time departures from $\Lambda$CDM \cite{Nunes:2022bhn, Chen:2025jnr}, underscoring the robustness and flexibility of this framework for testing beyond-standard cosmologies.

\section{Methodology}
\label{sec:methodology}

We divide the methodology section into three parts. First, we provide an overview of the analysis procedure, outlining the key steps of the computational pipeline used to obtain the results. Next, we describe the observational data sets employed to constrain the parameters of the cosmological models under consideration. Finally, we briefly discuss the statistical tests employed throughout the study.

\subsection{Analysis setup}

We implemented both parametrizations tested in this work as modifications to the background-level EoS within the Einstein–Boltzmann solver \texttt{CLASS-PT} \cite{Chudaykin:2020aoj}, a robust code that computes the nonlinear matter power spectra and biased tracer statistics within one-loop cosmological perturbation theory. The modified code was then interfaced with the \texttt{MontePython} \cite{Audren:2012wb, Brinckmann:2018cvx} sampler, which performs Monte Carlo analyses via Markov Chain Monte Carlo (MCMC) methods. In all analyses, convergence was ensured using the Gelman–Rubin criterion \cite{Gelman_1992}, with $R - 1 \leq 10^{-2}$.

The cosmological parameters sampled in this analysis include the physical baryon density $\omega_{\rm b} = \Omega_{\rm b} h^2$, the physical cold dark matter density $\omega_{\rm c} = \Omega_{\rm c} h^2$, the Hubble parameter today $H_0$, the amplitude of the primordial scalar power spectrum $A_{\mathrm{s}}$, and its spectral index $n_{\mathrm{s}}$. In addition to these standard parameters, we also sample the DE EoS parameters $w_0$ and $w_a$ in the context of the CPL parametrization. As an alternative approach, we also consider the parametrization $w_{\dagger}(a)$, characterized by the parameters $\Delta$ and $a_{\dagger}$. The flat prior ranges adopted are: 
\begin{align*}
    &\omega_{\rm b} \in [0.0, 1.0], \quad \omega_{\rm c} \in [0.0, 1.0], \quad H_0 \in [0.0, 100], \quad \ln(10^{10} A_s) \in [0.0, 10.0],\\
    &\quad w_0 \in [-1.2, 0.5], \quad w_a \in [-3.0, 1.0], \quad a_{\dagger} \in [0.25, 1.0], \quad \Delta \in [-1.0, 1.0].
\end{align*}
In the context of the EFTofLSS parameters, we also sampled a set of nuisance parameters associated with galaxy tracer bias. These include the linear bias $b_1$, the quadratic bias $b_2$, and the tidal bias $b_{G_2}$, each specified independently for the redshift bins considered. All post-processing of the Markov chains was performed using the \texttt{GetDist}\footnote{Available at \url{https://github.com/cmbant/getdist}.} package \cite{Lewis:2019xzd}, which we used to extract numerical results, including 1D posteriors and 2D marginalized probability contours.

\subsection{Data}

We constrain cosmological models using a combination of recent and complementary datasets, including BAO measurements from DESI DR2, FS power spectrum multipoles from BOSS DR12, and the latest compilations of Type Ia supernovae (SN Ia). Further details on each dataset are provided below. The three SN Ia samples are treated independently to enable robust cross-checks of their cosmological constraints.

\paragraph{Baryon Acoustic Oscillations (\texttt{DESI-DR2}).} This dataset includes observations of galaxies, quasars \cite{DESI:2025zgx}, and Lyman-$\alpha$ forest tracers \cite{DESI:2025zpo} from DESI Data Release 2. The corresponding measurements, summarized in Table IV of Ref. \cite{DESI:2025zgx}, provide isotropic and anisotropic BAO constraints over the redshift range $0.295 \leq z \leq 2.330$, divided into seven effective redshift bins. The BAO observables include the transverse comoving distance $D_{\mathrm{M}}/r_{\mathrm{d}}$, the Hubble distance $D_{\mathrm{H}}/r_{\mathrm{d}}$, and the volume-averaged distance $D_{\mathrm{V}}/r_{\mathrm{d}}$, all normalized by the sound horizon at the drag epoch, $r_{\mathrm{d}}$. We also account for correlations among these observables through the reported cross-correlation coefficients: $r_{\mathrm{V,M/H}}$, which encodes the correlation between $D_{\mathrm{V}}/r_{\mathrm{d}}$ and $D_{\mathrm{H}}/r_{\mathrm{d}}$, and $r_{\mathrm{M,H}}$, which quantifies the correlation between $D_{\mathrm{M}}/r_{\mathrm{d}}$ and $D_{\mathrm{H}}/r_{\mathrm{d}}$. We refer to this dataset as \texttt{DESI-DR2} throughout this work.

\paragraph{Full-shape power spectrum of BOSS DR12 (\texttt{FS}).} This dataset comprises FS measurements of the power spectrum monopole and quadrupole ($\ell = 0$ and $\ell = 2$, respectively) from the BOSS DR12 galaxy sample~\cite{alam2015eleventh}. The sample is divided into two effective redshift bins at $z_{\mathrm{eff}} = 0.38$ and $z_{\mathrm{eff}} = 0.61$, each further split into the Northern and Southern Galactic Caps (NGC and SGC), yielding four independent subsamples. This binning follows previous FS analyses~\cite{Philcox:2021kcw, BOSS:2016psr, Ivanov:2019pdj} and combines galaxies from the CMASS and LOWZ samples. We utilize the publicly available \texttt{CMASSLOWZTOT} catalog, together with the associated window functions and random catalogs, to accurately model survey geometry and selection effects. Covariance matrices used in the likelihood analysis are derived from 2048 \texttt{MultiDark-Patchy} mock catalogs~\cite{Kitaura:2015uqa, Rodriguez-Torres:2015vqa}, calibrated against $N$-body simulations. Throughout this work, we refer to this dataset combination as \texttt{FS}, encompassing the power spectrum multipoles (P$_0$/P$_2$), the real-space proxy (Q$_0$), and the BAO rescaling parameters (AP). We do not include measurements of the galaxy bispectrum multipoles in our analysis due to the considerably higher modeling complexity associated with these statistics. For further details on the likelihood construction, see~\cite{Philcox:2021kcw}.

\paragraph{PantheonPlus (\texttt{PP}).} The PantheonPlus sample~\cite{pantheonplus} consists of distance modulus measurements for 1701 light curves from 1550 distinct SN Ia events, spanning the redshift range $0.01 < z < 2.26$. We refer to this dataset as \texttt{PP}. 

\paragraph{Union 3.0 (\texttt{Union3}).} The Union 3.0 compilation comprises 2087 SN Ia events in the redshift range $0.05 < z < 2.262$, as presented in~\cite{Rubin:2023ovl}, with 1363 events overlapping with the PantheonPlus sample. This dataset adopts a fully Bayesian hierarchical modeling framework for systematic uncertainties, providing an alternative approach to SN cosmology. We refer to this dataset as \texttt{Union3}.

\paragraph{Dark Energy Survey Y5 (\texttt{DESY5}).} The final dataset corresponds to the Year 5 release of the Dark Energy Survey (DES), comprising a homogeneous sample of 1635 photometrically classified SN Ia over $0.1 < z < 1.3$, supplemented by 194 low-redshift SN Ia ($0.025 < z < 0.1$) shared with PantheonPlus~\cite{DES:2024tys}. We refer to this dataset as \texttt{DESY5}. It provides an independent test of SN cosmology based on a uniform and well-controlled photometric pipeline.

In order to constrain cosmological parameters without CMB data, we apply a Gaussian prior on the physical baryon density $\omega_{\rm b}=\Omega_{\rm b}h^2$, based on Big Bang Nucleosynthesis (BBN) predictions \cite{Consiglio:2017pot} and primordial deuterium and helium abundances \cite{Cooke:2017cwo, Aver:2015iza}, used in all combined analyses.

\subsection{Statistical tests}
To reinforce the robustness of our interpretations, we also perform a set of statistical and model selection tests, detailed below. As an initial diagnostic, we compute the difference in the minimum $\chi^2$ values between models, defined as $\Delta \chi^2_{\text{min}} \equiv \chi^2_{\text{min (MODEL)}} - \chi^2_{\text{min ($\Lambda$CDM)}}$. Since $\chi^2_{\text{min}} = -2 \ln \mathcal{L}_{\rm max}$, this provides a simple yet informative measure of relative goodness-of-fit under the assumption of equal prior volume. A negative $\Delta \chi^2_{\text{min}}$ indicates that the extended model achieves a better fit than $\Lambda$CDM, while a positive value indicates a worse fit.

In the context of bayesian model comparison, we adopt the following statistical tools to assess model performance and consistency:

\paragraph{Widely Applicable Information Criterion (WAIC).} To better evaluate model performance—accounting for both model complexity and parameter uncertainties—we adopt the WAIC~\cite{Watanabe:2010aic}, a Bayesian generalization of the well-known Akaike Information Criterion (AIC) \cite{Akaike:1974vps}. A key advantage of the WAIC for Bayesian analyses is that it is computed from the full posterior distribution rather than relying on point estimates. The WAIC is defined as
\begin{equation}
\mathrm{WAIC} = -2 \left[ \sum_{i=1}^{N_{\mathrm{data}}} \ln \left( \frac{1}{s} \sum_{n=1}^{s} \mathcal{L}_i(\theta^{(n)}) \right) 
- \sum_{i=1}^{N_{\mathrm{data}}} \mathrm{Var}_{n} \left( \ln \mathcal{L}_i(\theta^{(n)}) \right) \right],
\end{equation}
where $s$ is the number of posterior samples, $\mathcal{L}_i(\theta^{(n)})$ is the likelihood contribution of the $i$-th data point evaluated at the $n$-th posterior sample of the model parameters $\theta$, and the variance is computed over the posterior samples. The WAIC provides an asymptotically unbiased estimate of the model’s out-of-sample predictive accuracy, effectively balancing goodness-of-fit and model complexity. Models with lower WAIC values are preferred, as they exhibit superior expected predictive performance.

\paragraph{Bayesian Evidence (BE).} 
The BE offers a rigorous framework for model selection by integrating the likelihood over the full parameter space, thus naturally accounting for prior volume and model complexity. It defines the Bayes factor for models $\mathcal{M}_i$ and $\mathcal{M}_j$ as
\begin{equation}
\mathcal{B}_{ij} = \frac{p(\mathcal{M}_i|\mathbf{x})}{p(\mathcal{M}_j|\mathbf{x})} = \frac{\displaystyle \int d \boldsymbol{\theta}_i \, \pi(\boldsymbol{\theta}_i) \mathcal{L}(\mathbf{x} | \boldsymbol{\theta}_i)}{\displaystyle \int d \boldsymbol{\theta}_j \, \pi(\boldsymbol{\theta}_j) \mathcal{L}(\mathbf{x} | \boldsymbol{\theta}_j)},
\label{BayesFactor}
\end{equation}
where $\pi(\boldsymbol{\theta})$ denotes the prior and $\mathcal{L}$ the likelihood. Unlike WAIC, which targets predictive performance, the BE evaluates the relative support of the data for competing models while inherently penalizing complexity. We report $\ln \mathcal{B}_{ij}$ relative to $\Lambda$CDM, with positive (negative) values favoring $\Lambda$CDM (the extended model). Following~\cite{Kass:1995loi}, we classify $|\ln \mathcal{B}_{ij}|$ as:
\begin{equation}
|\ln \mathcal{B}_{ij}| =
\begin{cases}
< 1, & \text{Inconclusive}, \\
1-3, & \text{Positive evidence}, \\
3-5, & \text{Strong evidence}, \\
\geq 5, & \text{Very strong evidence}.\\
\end{cases}
\end{equation}
Since such fixed thresholds may overstate the significance of Bayes factors \cite{Keeley:2021dmx, Amendola:2024prl}, we also rely on complementary metrics and analyses throughout this work. We compute the Bayes factors using the publicly available \texttt{MCEvidence}, which evaluates the BE directly from MCMC posterior samples.

\section{Main Results}
\label{sec:results}
\begin{figure}[h!]
    \centering
    \includegraphics[width=0.43\textwidth]{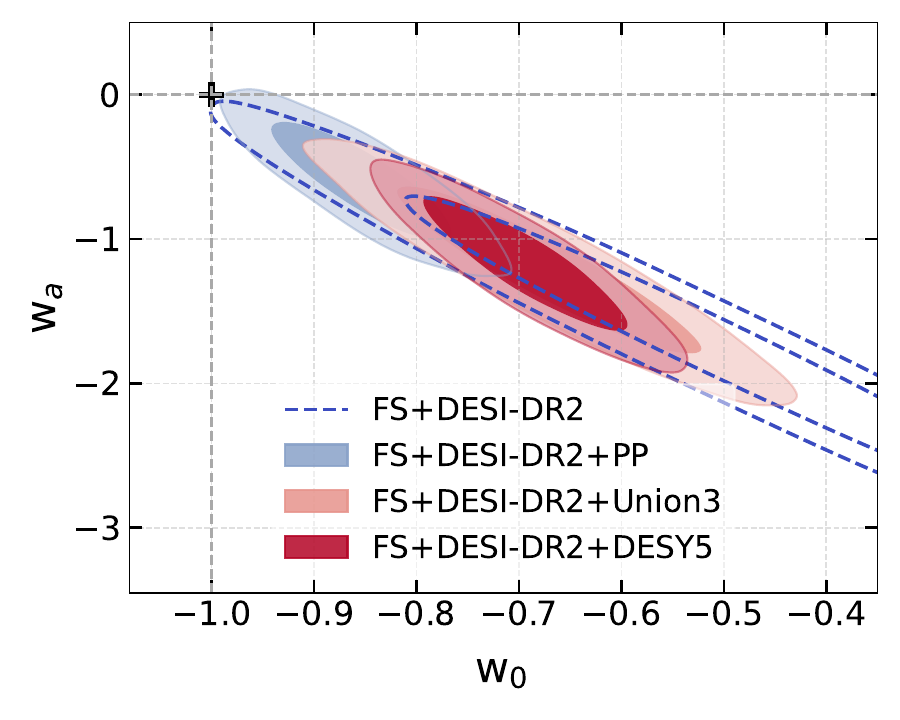}
    \includegraphics[width=0.43\textwidth]{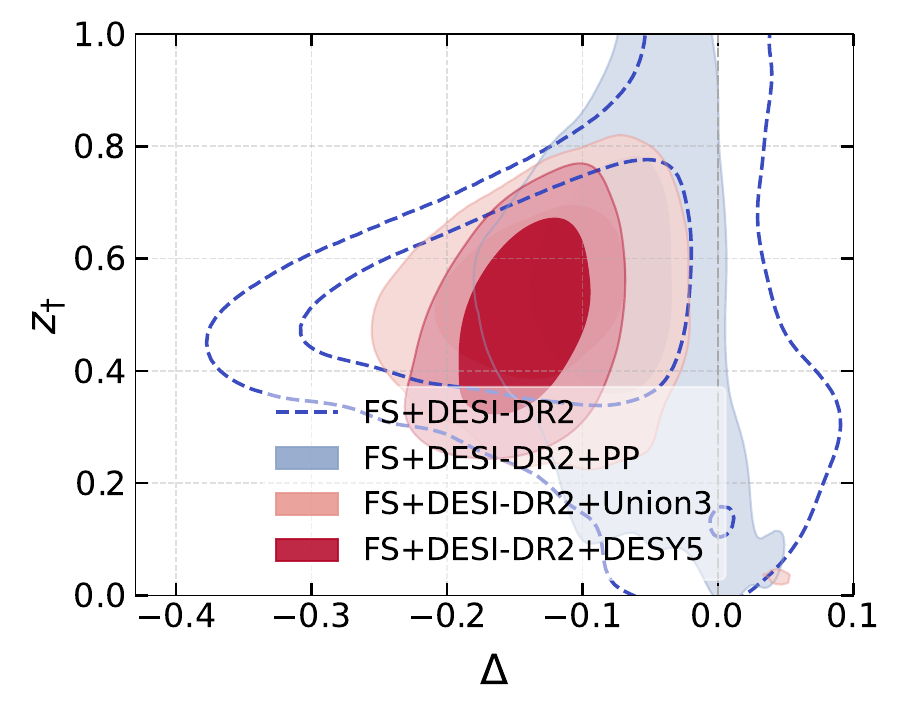}
    \caption{Left panel: 68\% and 95\% confidence contours for $w_a$ and $w_0$ from different data sets (see the legend) in the $w_0 w_a$CDM model. Right panel: Same contours for $\Delta$ and $z_{\dagger}$ in the $w_{\dagger}$CDM model.}
    \label{fig:contournsDDE}
\end{figure}

The main synthesis of our results, which will be discussed throughout this section, is presented in figure~\ref{fig:contournsDDE}, which illustrates the behavior of the parameters characterizing the functional form of the EoS, both showing indications of deviations from the $\Lambda$CDM model.
We provide a detailed discussion of the derived constraints in two subsections: the first focusing on the $(w_0, w_a)$ parametrization, and the second on the $w_{\dag}$ parametrization. In both cases, we introduce additional constrained parameters and examine the resulting evidence, as well as potential tensions with recent results in the literature.

\subsection{$w_{0}w_{a}$CDM}

Table~\ref{table:CPL} summarizes the main statistical results obtained in our analyses for the CPL model. For reference, the corresponding results for the standard $\Lambda$CDM model are provided in appendix~\ref{sec:appendixa}.

\begin{table*}[htpb!]
\centering
\Huge
\renewcommand{\arraystretch}{1.8}
\resizebox{\textwidth}{!}{
\begin{tabular}{|l|c|c|c|c|}
\hline
\textbf{Parameter} & \textbf{FS+DESI-DR2} & \textbf{FS+DESI-DR2+PP} & \textbf{FS+DESI-DR2+Union3} & \textbf{FS+DESI-DR2+DESY5} \\ 
\hline
$10^{2} \Omega_{\rm b} h^2$ & $2.271\pm 0.036$ ($2.271^{+0.074}_{-0.070}$) & $2.270\pm 0.037$ ($2.270^{+0.075}_{-0.072}$) & $2.272\pm 0.037$ ($2.272^{+0.076}_{-0.070}$) & $2.273\pm 0.038$ ($2.273^{+0.077}_{-0.074}$)  \\
$\Omega_{\rm c} h^2$ & $0.1309^{+0.0063}_{-0.0055}$ ($0.131^{+0.011}_{-0.013}$) & $0.1254\pm 0.0059$ ($0.125^{+0.011}_{-0.011}$) & $0.1291\pm 0.0058$ ($0.129^{+0.011}_{-0.012}$) & $0.1295\pm 0.0057$ ($0.130^{+0.011}_{-0.011}$)  \\
$n_{s}$ & $0.894\pm 0.049$ ($0.894^{+0.096}_{-0.096}$) & $0.925\pm 0.047$ ($0.925^{+0.092}_{-0.095}$) & $0.904\pm 0.047$ ($0.904^{+0.091}_{-0.096}$) & $0.902\pm 0.048$ ($0.902^{+0.093}_{-0.093}$)  \\
$\ln(10^{10} A_{s})$ & $2.76\pm 0.13$ ($2.76^{+0.26}_{-0.25}$) & $2.79\pm 0.15$ ($2.79^{+0.28}_{-0.29}$) & $2.76\pm 0.14$ ($2.76^{+0.27}_{-0.27}$) & $2.74\pm 0.14$ ($2.74^{+0.27}_{-0.27}$)  \\
$\Omega_{\mathrm{m}}$ & $0.354^{+0.024}_{-0.020}$ ($0.354^{+0.040}_{-0.042}$) & $0.3181\pm 0.0082$ ($0.318^{+0.016}_{-0.016}$) & $0.338\pm 0.012$ ($0.338^{+0.023}_{-0.024}$) & $0.3354\pm 0.0086$ ($0.335^{+0.017}_{-0.017}$)  \\
$H_0 \, [\mathrm{km/s/Mpc}]$ & $66.1^{+1.4}_{-1.7}$ ($66.1^{+3.2}_{-2.8}$) & $68.38\pm 0.91$ ($68.4^{+1.8}_{-1.7}$) & $67.20\pm 0.96$ ($67.2^{+1.9}_{-1.9}$) & $67.52\pm 0.85$ ($67.5^{+1.7}_{-1.6}$)  \\
$S_8$ & $0.775\pm 0.047$ ($0.775^{+0.097}_{-0.090}$) & $0.749\pm 0.045$ ($0.749^{+0.091}_{-0.087}$) & $0.761\pm 0.045$ ($0.761^{+0.091}_{-0.084}$) & $0.758\pm 0.045$ ($0.758^{+0.090}_{-0.085}$)  \\
$\mathrm{w}_0$ & $-0.52^{+0.22}_{-0.18}$ ($-0.52^{+0.36}_{-0.37}$) & $-0.852\pm 0.057$ ($-0.85^{+0.11}_{-0.11}$) & $-0.670\pm 0.097$ ($-0.67^{+0.19}_{-0.19}$) & $-0.692\pm 0.064$ ($-0.69^{+0.13}_{-0.12}$)  \\
$\mathrm{w}_a$ & $-1.70^{+0.63}_{-0.74}$ ($-1.7^{+1.2}_{-1.2}$) & $-0.60\pm 0.27$ ($-0.60^{+0.51}_{-0.53}$) & $-1.23\pm 0.38$ ($-1.23^{+0.74}_{-0.75}$) & $-1.18\pm 0.30$ ($-1.18^{+0.58}_{-0.60}$)  \\
\hline
$\Delta \chi^{2}_{\mathrm{min}}$ & -5.50 & -6.30 & -10.97 & -25.58 \\
$\Delta \mathrm{WAIC}$ & -2.78 & -3.34 & -9.14 & -23.41 \\
$\ln \mathcal B_{ij}$ & 0.76 & 1.61 & -1.75 & -8.72 \\
\hline
\end{tabular}}
\caption{$68\%$ and $95\%$ confidence level (CL) constraints for the parameters of the CPL model. In the last rows, we report the quantities $\Delta \chi^2_{\text{min}} \equiv \chi^2_{\text{min (CPL)}} - \chi^2_{\text{min ($\Lambda$CDM)}}$, $\Delta _\text{WAIC} \equiv \text{WAIC}_{\text{CPL}} - \text{WAIC}_{\text{$\Lambda$CDM}}$, and $\ln \mathcal{B}{ij}$ (the natural logarithm of the Bayes factor) to compare the models' statistical performance. Negative values of $\Delta \chi^2_{\text{min}}$, $\Delta \text{WAIC}$, and $\ln \mathcal{B}_{ij}$ favor the CPL model over $\Lambda$CDM, while positive values indicate preference for $\Lambda$CDM.}
\label{table:CPL}
\end{table*}

Our first set of analyses was performed using the FS+DESI-DR2 dataset. We observe that the constraints on the primordial power spectrum parameters, namely $A_s$ and $n_s$, are slightly lower when compared to the corresponding $\Lambda$CDM Planck-CMB constraints. As discussed in the previous section and quantitatively illustrated in figure \ref{fig:contournsDDE}, the introduction of evolving DE parameters influences mainly the amplitude of the galaxy power spectrum. Consequently, these models exhibit enhanced sensitivity to variations in parameters such as $A_s$ and  $\Omega_m$, which is reflected in the shifts observed in the joint FS+DESI-DR2 analysis, when compared with Planck-CMB values. The obtained constraints on the parameters $\Omega_b$ and $H_0$ are consistent with their canonical values in the $\Lambda$CDM model as inferred from Planck-CMB data.

As a direct consequence of the shifts in the baseline values of $A_s$ and $\Omega_m$, we find a reduced amplitude of matter fluctuations, with $S_8 = 0.775 \pm 0.047$, notably lower than the value inferred from Planck-CMB observations. This also reinforces a well-documented tendency in the literature for BOSS data to prefer smaller values of this parameter \cite{Ivanov:2024xgb, Philcox:2021kcw, Semenaite:2022unt, Carrilho:2022mon}. Another noteworthy result is that the dataset exhibits a clear statistical preference for a DE evolution characterized by $w_0 > -1$ and $w_a < 0$, a trend primarily driven by the inclusion of BAO measurements from the DESI DR2 samples. These findings point to a mild preference for a phantom-crossing behavior of the DE EoS at late times. This interpretation is further supported by improvements in model comparison metrics, with $\Delta\mathrm{WAIC} = -2.78$ and a minimum $\chi^2$ difference of $\Delta\chi^2_{\mathrm{min}} = -5.50$, both favoring the CPL model over $\Lambda$CDM. However, we note that the correlation between FS data and DESI-DR2 measurements is not included in our analyses, since the corresponding covariance matrices are not publicly available. This caveat should be kept in mind when interpreting the model comparison statistics, in particular the Bayes factor, for which we obtain $\ln B_{ij} = 0.76$. This value suggests that the evidence does not point to a decisive preference over $\Lambda$CDM.

In what follows, we consider the impact of adding three different SNIa samples to the analysis. We begin by examining the combined dataset FS+DESI-DR2+PP, where a slight improvement is observed in the main cosmological parameters of interest within the baseline model. The same statistical behavior previously identified in the joint FS+DESI-DR2 analysis persists, but with a noticeable reduction in the associated error bars, indicating improved precision. Additionally, we observe a shift in the best-fit value of the matter density parameter, now trending toward $\Omega_m \sim 0.31$ with the inclusion of the PP sample, compared to the higher estimate of $\Omega_m \sim 0.35$ obtained from FS+DESI-DR2 alone. However, it is important to emphasize that all estimates remain fully consistent within the respective uncertainties. Therefore, this shift does not indicate any tension between the datasets, but rather reflects a mild change in the preferred value.

The constraints on the parameter $S_8$ continue to favor relatively low values, with $S_8 = 0.749 \pm 0.045$, aligning with trends seen in previous analysis. In the $w_0$-$w_a$ parameter space, we continue to find the characteristic pattern with $w_a < 0$ and $w_0 > -1$, now with an evident improvement in statistical significance. Despite the tighter constraints in parameter space, the Bayesian model comparison yields $\ln \mathcal{B}_{ij} = 1.61$, corresponding to positive evidence in favor of $\Lambda$CDM. This is notably due to the substantially larger data volume in the FS+DESI-DR2+PP combination compared to FS+DESI-DR2 alone. This difference reflects how the inclusion of additional measurements increases the statistical weight, leading to stronger constraints that tend to favor the simpler $\Lambda$CDM model by penalizing the extra flexibility of the CPL parameterization through the Bayesian evidence calculation.

In what follows, we consider the joint analyses FS+DESI-DR2+Union3 and FS+DESI-DR2+DESY5. These combinations provide the strongest observational constraints among all the tested configurations. The full cosmological parameter space is more tightly constrained, as shown in Table~\ref{table:CPL}. In particular, we highlight the results for $S_8$, which are $S_8 = 0.761 \pm 0.045$ for FS+DESI-DR2+Union3 and $S_8 = 0.758 \pm 0.045$ for FS+DESI-DR2+DESY5, both of which are consistent with previous analyses that indicate slightly lower values compared to Planck-based constraints.

As observed in the other cases, the $w_0$--$w_a$ parameter space remains well constrained and exhibits a strong tendency toward a phantom crossing behavior. These two final joint analyses also suggest significant deviations from the standard $\Lambda$CDM model. Specifically, we obtain $\Delta \mathrm{WAIC} = -9.14$ for FS+DESI-DR2+Union3 and $\Delta \mathrm{WAIC} = -23.41$ for FS+DESI-DR2+DESY5. As previously discussed, the WAIC criterion strongly penalizes models with additional parameters; nevertheless, we find compelling statistical support for the CPL model in both cases. When adopting an even more robust approach based on the Bayes factor—which accounts for the full shape of the posterior—we find $\ln \mathcal{B}_{ij} = -1.75$ and $\ln \mathcal{B}_{ij} = -8.72$ for FS+DESI-DR2+Union3 and FS+DESI-DR2+DESY5, respectively. The latter represents very strong evidence in favor of the CPL model. In terms of a Gaussian-equivalent interpretation, this corresponds to a deviation of approximately $5\sigma$.

For a clearer understanding of our results, in Appendix \ref{sec:appendixb} we assess the specific role played by the FS data in constraining DE parameters. In particular, we perform complementary analyses in which the FS data are excluded, allowing us to quantify the additional information provided by FS beyond that already captured by BAO and supernova observations. A discussion of these comparisons is presented there.

\begin{figure}[htpb!]
    \centering
    \includegraphics[width=\columnwidth]{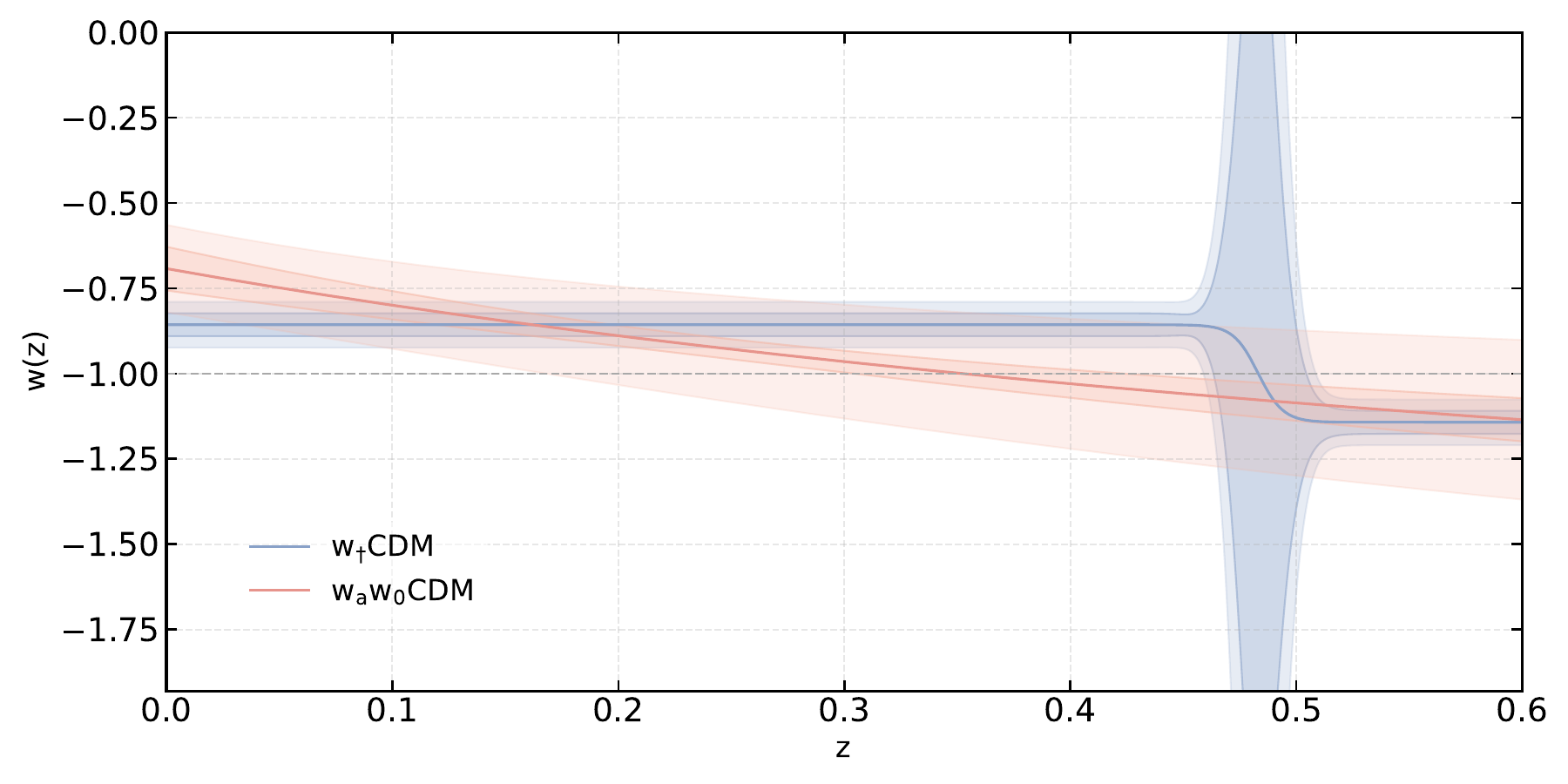} 
    \caption{Statistical reconstruction of the DE EoS for the $w_0w_a$ and $w_{\dag}$ parametrizations, showing 1$\sigma$ and 2$\sigma$ confidence levels from the combined FS+DESI+DESY5 dataset analyses.}
    \label{fig:reconstruction}
\end{figure}

Figure~\ref{fig:contournsDDE} (left panel) shows the 2D confidence contours at 68\% and 95\% confidence levels for the parameters $w_a$ and $w_0$, derived from all the joint analyses discussed above. A clear trend emerges in the parameter space, with a strong statistical preference for $w_a < 0$ and $w_0 > -1$, indicating a deviation from the standard $\Lambda$CDM behavior. Figure~\ref{fig:reconstruction} presents the statistical reconstruction of the DE EoS $w(z)$ for the CPL model. Notably, the reconstruction reveals the presence of a phantom crossing at low redshift, occurring around $z_{\rm transition} \sim 0.38$, based on the best-fit values. Therefore, our main results for this case provide strong statistical evidence for a phantom crossing at late times, independently of CMB primary anisotropy data.

A very recent and particularly interesting study to compare with our results is \cite{Chen:2025jnr}, although their focus lies on exploring simulation-based priors and they also include CMB data. When comparing our results to theirs under conservative priors using CMB+FS+PP+DESI-DR2, we find that our values of $\Omega_m$ are compatible within the uncertainties. For $H_0$, we obtain consistent results in our FS+DESI-DR2+Union3 and FS+DESI-DR2+DESY5 combinations, while FS+DESI-DR2 and FS+DESI-DR2+PP yield slightly lower and slightly higher values, respectively, although without statistical significance. Regarding the CPL parameters, we find notably smaller values of $w_a$ and slightly higher values of $w_0$ (except for FS+DESI-DR2+PP, where $w_0$ remains consistent). This behavior is most likely a direct consequence of the inclusion of CMB data, which strongly constrains the late-time evolution of the EoS. Finally, it is important to note that the error bars reported in \cite{Chen:2025jnr} are significantly smaller, which is naturally attributed both to the inclusion of Planck data and to the fact that their FS analysis incorporates additional multipole and bispectrum information.

A complementary methodological reference is provided by \cite{DESI:2024hhd}, which presents cosmological constraints from FS modeling of DESI clustering measurements. Unlike BOSS, the DESI analysis employs a blinding procedure to mitigate experimenter bias. Beyond this difference between the pipelines, the DESI FS+BAO likelihood adopts a higher-dimensional parameter space, including bias coefficients, counterterms, and stochastic terms per redshift bin, all jointly constrained with cosmological parameters using the full covariance between FS and BAO observables. This contrasts with the more compact BOSS DR12 likelihood, based on pre-validated monopole and quadrupole measurements with fewer nuisance parameters and fixed redshift bins \cite{Ivanov:2019pdj, Ivanov:2021fbu, Philcox:2021kcw, Chudaykin:2020aoj}, reflecting the data precision and modeling practices of its time.\footnote{or a comprehensive understanding of the different methodological frameworks, we recommend that readers consult the original analyses presented in \cite{DESI:2024hhd} and \cite{Philcox:2021kcw}.} We emphasize that comparing internal pipeline differences across the methodologies of large collaborations teams lies beyond the scope of the present work. In the future, we plan to use the pre-reconstruction galaxy power spectrum multipoles from public DESI data for FS analysis once they become available, which will enable a new statistical comparison.

In terms of parameter constraints, our results are broadly consistent with those of \cite{DESI:2024hhd}, with a few notable differences. For the matter density, we find $\Omega_{\mathrm{m}}=0.3181\pm0.0082$ (FS+DESI-DR2+PP), in excellent agreement with their DESI-DR1(FS+BAO)+PP value of $0.3084\pm0.0089$. Our FS+DESI-DR2+Union3 result, $0.338\pm0.012$, is slightly higher but remains consistent within $2\sigma$. Similarly, our Hubble constant determinations overlap closely: for example, $H_0 = 68.38\pm0.91$ km s$^{-1}$Mpc$^{-1}$ (FS+DESI-DR2+PP) matches their $68.4\pm1.1$ km s$^{-1}$Mpc$^{-1}$ (DESI-DR1(FS+BAO)+PP). For the clustering amplitude, our $S_8$ values ($\sim0.75$–$0.78$) are slightly lower than theirs ($\sim0.81$–$0.82$), consistent with the trend generally observed for BOSS data. The largest differences appear in the DE equation of state: our FS+DESI-DR2+PP constraint of $w_0=-0.852\pm0.057$ is in close agreement with their $w_0=-0.875\pm0.072$, whereas our FS+DESI-DR2+Union3 and FS+DESI-DR2+DESY5 combinations yield mildly higher values ($w_0\approx-0.67$). Likewise, our $w_a$ determinations tend toward more negative central values (e.g., $-1.23\pm0.38$ for FS+DESI-DR2+Union3) compared with their range of roughly $-0.61$ to $-1.12$, although uncertainties overlap. Overall, these comparisons confirm that our analysis reproduces the main cosmological constraints obtained by the DESI FS pipeline, while naturally reflecting the impact of distinct data combinations and methodological choices.

\subsection{$w_{\dagger}$CDM framework}

For the $w_{\dag}$ model, table~\ref{table:wdag} summarizes the main statistical results obtained in our analyses. For reference, the corresponding results for the standard $\Lambda$CDM model are provided in Appendix~\ref{sec:appendixa}.

\begin{table*}[htpb!]
\centering
\Huge
\renewcommand{\arraystretch}{1.8}
\resizebox{\textwidth}{!}{
\begin{tabular}{|l|c|c|c|c|}
\hline
\textbf{Parameter} & \textbf{FS+DESI-DR2} & \textbf{FS+DESI-DR2+PP}  & \textbf{FS+DESI-DR2+Union3} & \textbf{FS+DESI-DR2+DESY5} \\ \hline
$10^{2} \Omega_{\rm b} h^2$ & $2.269\pm 0.037$ ($2.269^{+0.074}_{-0.071}$) & $2.271\pm 0.038$ ($2.271^{+0.079}_{-0.072}$) & $2.271\pm 0.037$ ($2.271^{+0.076}_{-0.072}$) & $2.271\pm 0.037$ ($2.271^{+0.074}_{-0.071}$)  \\
$\Omega_{\rm c} h^2$ & $0.1209^{+0.0042}_{-0.0048}$ ($0.1209^{+0.0095}_{-0.0086}$) & $0.1191^{+0.0035}_{-0.0040}$ ($0.1191^{+0.0080}_{-0.0073}$) & $0.1210^{+0.0038}_{-0.0044}$ ($0.1210^{+0.0089}_{-0.0078}$) & $0.1226^{+0.0041}_{-0.0055}$ ($0.1226^{+0.0098}_{-0.0090}$)  \\
$n_{s }$ & $0.944\pm 0.046$ ($0.944^{+0.087}_{-0.091}$) & $0.954\pm 0.043$ ($0.954^{+0.081}_{-0.088}$) & $0.943\pm 0.045$ ($0.943^{+0.086}_{-0.090}$) & $0.934\pm 0.045$ ($0.934^{+0.088}_{-0.090}$)  \\
$\ln(10^{10} A_{s})$ & $2.90\pm 0.12$ ($2.90^{+0.24}_{-0.24}$) & $2.91\pm 0.12$ ($2.91^{+0.24}_{-0.24}$) & $2.90\pm 0.13$ ($2.90^{+0.25}_{-0.24}$) & $2.87\pm 0.14$ ($2.87^{+0.26}_{-0.27}$)  \\
$\Omega_{\rm m}$ & $0.319^{+0.014}_{-0.020}$ ($0.319^{+0.032}_{-0.029}$) & $0.3115\pm 0.0071$ ($0.311^{+0.014}_{-0.014}$) & $0.3217\pm 0.0094$ ($0.322^{+0.019}_{-0.019}$) & $0.3246\pm 0.0075$ ($0.325^{+0.015}_{-0.014}$)  \\
$H_0 \, [\mathrm{km/s/Mpc}]$ & $67.2\pm 1.3$ ($67.2^{+2.6}_{-2.4}$) & $67.62^{+0.69}_{-0.82}$ ($67.6^{+1.6}_{-1.4}$) & $66.99^{+0.79}_{-0.90}$ ($67.0^{+1.8}_{-1.6}$) & $67.06^{+0.71}_{-0.83}$ ($67.1^{+1.6}_{-1.4}$)  \\
$S_8$ & $0.770\pm 0.048$ ($0.770^{+0.099}_{-0.091}$) & $0.763^{+0.042}_{-0.047}$ ($0.763^{+0.090}_{-0.082}$) & $0.773^{+0.042}_{-0.049}$ ($0.773^{+0.095}_{-0.085}$) & $0.769\pm 0.046$ ($0.769^{+0.093}_{-0.086}$)  \\
$z_{\dag}$ & $0.625^{+0.051}_{-0.24}$ ($0.62^{+1.2}_{-0.62}$) & $0.633^{+0.077}_{-0.21}$ ($0.63^{+0.67}_{-0.61}$) & $0.532^{+0.11}_{-0.095}$ ($0.53^{+0.22}_{-0.27}$) & $0.49\pm 0.11$ ($0.49^{+0.21}_{-0.21}$)  \\
$\Delta$ & $-0.113\pm 0.088$ ($-0.11^{+0.17}_{-0.17}$) & $-0.075^{+0.031}_{-0.036}$ ($-0.075^{+0.075}_{-0.065}$) & $-0.129^{+0.043}_{-0.048}$ ($-0.129^{+0.095}_{-0.096}$) & $-0.143^{+0.036}_{-0.031}$ ($-0.143^{+0.062}_{-0.069}$)  \\
\hline
$\Delta \chi^{2}_{\mathrm{min}}$ & -5.09 & -8.18 & -11.32 & -21.12 \\
$\Delta \mathrm{WAIC}$ & -1.30 & -2.80 & -7.67 & -19.29 \\
$\ln \mathcal B_{ij}$ & 1.62 & 0.78 & -0.73 & -6.65 \\
\hline
\end{tabular}}
\caption{Same as in table \ref{table:CPL}, but for the $w_{\dag}$CDM model.}
\label{table:wdag}
\end{table*}

Starting from the baseline FS+DESI-DR2 case, we find that the estimates for the matter density and the Hubble parameter are largely consistent with those derived from Planck-CMB observations. However, when examining the amplitude of matter fluctuations, we obtain $S_8 = 0.770 \pm 0.048$, which is notably lower than the corresponding values reported by Planck-CMB. This highlights a persistent tendency for LSS data to favor smaller values of $S_8$ compared to early-universe probes.

Adding supernova samples sequentially improves the precision of key cosmological parameters. With the inclusion of the PP sample, we find $\Omega_m = 0.3115\pm 0.0071$ and $S_8 = 0.763^{+0.042}_{-0.047}$, reflecting a modest downward shift in the matter density alongside tighter constraints on the clustering amplitude. The addition of Union3 or DES-Y5 samples further refines these estimates. For FS+DESI-DR2+Union3, we obtain $S_8 = 0.773^{+0.042}_{-0.049}$, whereas FS+DESI-DR2+DESY5 yields $S_8 = 0.769\pm 0.046$. In all cases, the results support relatively low values of $S_8$, consistent with earlier indications of mild tension with Planck-derived $\Lambda$CDM expectations.

\begin{figure}[htpb!]
    \centering
    \includegraphics[width=0.9\columnwidth]{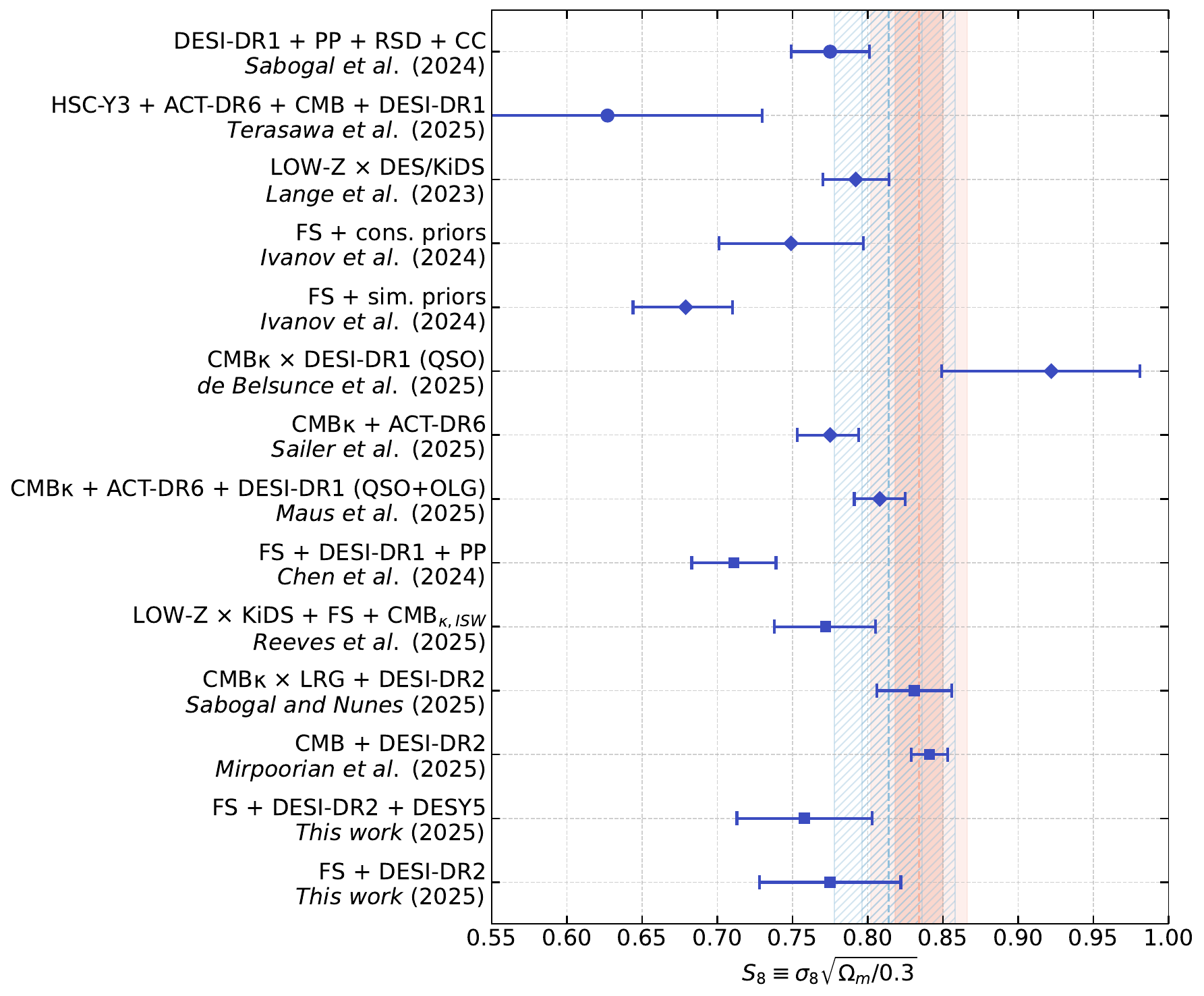} 
    \caption{Whisker plot displaying 68\% confidence intervals on $S_8$, as inferred from a range of recent analyses under different cosmological assumptions. Square markers indicate results obtained using the CPL parameterization: $S_8 = 0.775^{+0.047}_{-0.047}$ from FS+DESI-DR2 [\textit{This work}], $S_8 = 0.758^{+0.045}_{-0.045}$ from FS+DESI-DR2+DESY5 [\textit{This work}], $S_8 =0.711^{+0.028}_{-0.028}$ from FS+DESI-DR1+PP \cite{Chen:2024vuf}, $S_8 =0.841^{+0.012}_{-0.012}$ from CMB+DESI-DR2 \cite{Mirpoorian:2025rfp}, $S_8 =0.831^{+0.025}_{-0.025}$ from CMB$\kappa\times$LRG+DESI-DR2 \cite{Sabogal:2025jbo}, and $S_8 =0.772^{+0.034}_{-0.034}$ from LOW-Z$\times$(KiDS+FS+CMB$_{\kappa,\mathrm{ISW}}$) \cite{Reeves:2025axp}. Diamond markers correspond to standard $\Lambda$CDM analyses: $S_8 =0.679^{+0.035}_{-0.031}$ and $S_8 =0.749^{+0.048}_{-0.048}$ from FS with simulation and conservative priors \cite{Ivanov:2024xgb}, $S_8 =0.808^{+0.017}_{-0.017}$ from CMB$\kappa$+ACT-DR6+DESI-DR1 (QSO+OLG) \cite{Maus:2025rvz}, $S_8 =0.775^{+0.022}_{-0.022}$ from CMB$\kappa$+ACT-DR6 \cite{Sailer:2024jrx}, $S_8 =0.922^{+0.059}_{-0.073}$ from CMB$\kappa\times$DESI-DR1 (QSO) \cite{deBelsunce:2025qku}, and $S_8 =0.792^{+0.022}_{-0.022}$ from LOW-Z$\times$DES/KiDS \cite{Lange:2023khv}. Circular markers denote alternative scenarios, yielding $S_8 =0.627^{+0.103}_{-0.127}$ from HSC-Y3+ACT-DR6+CMB+DESI-DR1 \cite{Terasawa:2025fpf} and $S_8 =0.775^{+0.026}_{-0.026}$ from DESI-DR1+PP+RSD+CC \cite{Sabogal:2024yha}. The salmon and hatched light blue bands indicate the $1\sigma$ and $2\sigma$ confidence intervals from \textit{Planck} CMB and the latest KiDS results, respectively.}
\label{S8_plot}
\end{figure}

Figure~\ref{S8_plot} shows a whisker plot comparing our inferred $S_8$ values with recent estimates from diverse cosmological probes. These measurements rely on specific assumptions about the matter power spectrum, making their comparison particularly informative. In the plot, square markers indicate CPL results, diamonds correspond to standard $\Lambda$CDM analyses, and circles represent alternative scenarios such as modified gravity or dark sector interactions. Our values—$S_8=0.775\pm0.047$ (FS+DESI-DR2) and $0.758\pm0.045$ (FS+DESI-DR2+DESY5)—agree well with other FS-based studies like \cite{Chen:2024vuf} and \cite{Reeves:2025axp}, though they are slightly above the conservative estimates from \cite{Ivanov:2024xgb}. By contrast, they lie noticeably below analyses including CMB temperature and lensing data, such as \cite{Mirpoorian:2025rfp}, \cite{Sabogal:2025jbo}, and \cite{deBelsunce:2025qku}, reflecting the typical trend of late-time probes favoring lower $S_8$ relative to early-universe constraints.

Focusing on $\Lambda$CDM (highlighted by diamond markers), we observe the familiar pattern: FS or weak lensing analyses without direct CMB input tend to yield lower $S_8$ values, often consistent with our results, while those combining CMB data generally report higher $S_8$, accentuating the tension. Circular markers represent alternative scenarios, where \cite{Terasawa:2025fpf} constrains models with modified gravity and \cite{Sabogal:2024yha} investigates dark sector interactions—both approaches that can either ease or reproduce the observed discrepancy depending on the underlying physics. Overall, our estimates remain on the lower side of the current range, reinforcing the interpretation that the $S_8$ tension largely reflects differences between late-time structure growth and early-universe measurements.

Examining the DE sector, we consistently find evidence for a phantom-crossing-like behavior, with the parameter $\Delta$ remaining negative across all dataset combinations and becoming increasingly well constrained as additional supernova samples are incorporated, reaching $\Delta = -0.143^{+0.036}_{-0.031}$ in the FS+DESI-DR2+DESY5 case. This trend is accompanied by relatively low transition redshifts $z_{\dag}$, attaining $0.49\pm 0.11$ in the most stringent configuration. In parallel, we observe a progressive statistical preference for the $w_{\dag}$CDM model over $\Lambda$CDM as more data are included, reflected both in the improvements of the minimum $\chi^2$ and in the WAIC differences, culminating in $\Delta\mathrm{WAIC} = -19.29$ for FS+DESI-DR2+DESY5. The Bayes factor analysis similarly indicates growing support for the extended model, yielding $\ln \mathcal{B}_{ij} = -6.65$ for the same combination—classified as very strong evidence under standard interpretative scales. Taken together, these findings underscore the robustness of the inferred deviations from standard $\Lambda$CDM dynamics when probed through the $w_{\dag}$CDM parameterization.

An important theoretical aspect of our findings is the indication of a possible crossing of the phantom divide, $w=-1$, in the reconstructed dark-energy equation of state. Such behavior is known to challenge the consistency of simple single-field quintessence models, since a canonical scalar field cannot stably cross the $w=-1$ boundary without introducing pathologies such as ghost instabilities or violations of the null energy condition \cite{Carroll:2003st, Caldwell:2003vq}. Several theoretical frameworks have been proposed to accommodate this phenomenon, including quintom scenarios with two scalar fields, k-essence models, or effective field theory descriptions that remain well-behaved near the crossing \cite{Vikman:2004dc, Cai:2009zp, Hu:2007nk, Hu:2004kh, Deffayet:2010qz, Deffayet:2025lnj}. Comprehensive discussions of these possibilities and their stability conditions can be found in the review \cite{Copeland:2006wr} and related analyses \cite{Cline:2003gs, Guo:2004fq}.

\begin{figure}[htpb!]
    \centering
    \includegraphics[width=\columnwidth]{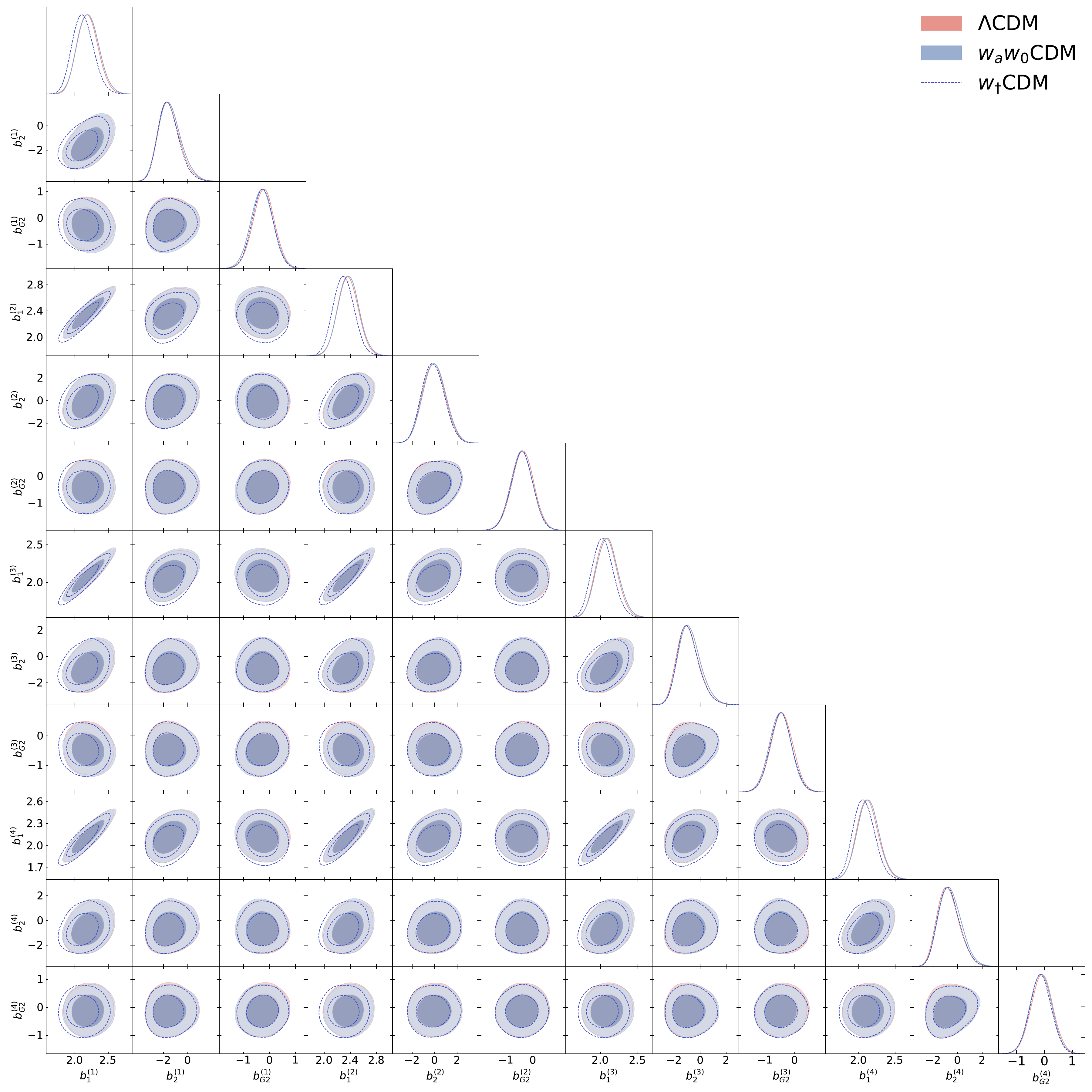} 
    \caption{1D posteriors and 2D marginalized probability contours for the bias parameters $b_1^{(i)}$, $b_2^{(i)}$, and $b_{G_2}^{(i)}$ ($i=1,\ldots,4$), corresponding to the four independent subsamples defined by redshift and sky region (LOWZ/CMASS in NGC/SGC). Results are shown for all cosmological models considered in this work, as indicated in the legend.}
    \label{bias_models}
\end{figure}

To conclude our discussion, figure \ref{bias_models} displays the one-dimensional posterior distributions and the two-dimensional marginalized confidence regions (at the 68\% and 95\% confidence levels) for the galaxy bias parameters $b_1$, $b_2$, and $b_{G2}$, derived from the FS dataset at effective redshifts $z = 0.38$ and $0.61$. As shown, all bias parameters are statistically consistent across the different cosmological models considered in this analysis. This agreement indicates that the adopted bias expansion scheme, as well as the perturbative kernels underlying the theoretical framework, remain robust and largely model-independent within the precision of current data and the theoretical assumptions employed, as discussed in more detail in Section \ref{EFT}. Consequently, we find no evidence of any statistical systematics in the estimation of the bias parameters arising from the use of different underlying cosmological frameworks.

\section{Final Remarks}
\label{sec:finalremarks}
In this work, we present a robust investigation of observational constraints on the evolution of the DE EoS using low-redshift cosmological probes. Our primary focus is on the impact of DE dynamics and its sensitivity to publicly available data from the FS galaxy power spectrum from the BOSS DR12 sample. For this purpose, we initially employed the traditional CPL parameterization in a first round of analyses. Subsequently, we explored a novel approach proposed in \cite{Scherer:2025esj}, which provides a minimal framework to investigate possible late-time transitions in the DE sector, independent of theoretical priors and allowing for a flexible reconstruction of the EoS as a function of redshift.

We demonstrate that even without including CMB data, FS measurements, when combined with recent BAO data from DESI and SNIa compilations, constitute a powerful cosmological probe. These combinations yield statistical precision sufficient to constrain extended cosmological models with two additional parameters beyond $\Lambda$CDM. Across all analyses performed, we find a strong statistical indication of a phantom crossing behavior in the redshift interval $z \sim 0.4$--$0.5$. Through multiple statistical metrics, joint analyses such as FS+DESI-DR2+Union3 and FS+DESI-DR2+DESY5 provide strong evidence in favor of a phantom crossing, excluding the standard $\Lambda$CDM model at up to $5\sigma$ confidence in some cases (e.g., FS+DESI-DR2+DESY5). Our results are in agreement with other recent independent studies that also report significant evidence for a transition across the phantom divide line \cite{Lee:2025pzo, Ozulker:2025ehg, Nojiri:2025low, Gonzalez-Fuentes:2025lei, Tiwari:2024gzo, Cai:2025mas, Wang:2025hlh, Bedroya:2025fwh}. It is particularly important to emphasize that our findings are independent of CMB anisotropy measurements, highlighting the strength of FS data as a competitive observational tool for probing extended cosmological parameter spaces beyond $\Lambda$CDM.

In light of current cosmological tensions related to $S_8$ and $H_0$, our main findings can be summarized as follows:
\begin{itemize}
 \item Within this extended parameter space and under the phantom crossing trend, we observe a consistent preference for lower values of $S_8$ when compared to constraints derived from primary CMB anisotropy measurements. Our results are in agreement with other FS-based analyses performed within the $\Lambda$CDM framework. Although we do observe a tendency toward lower $S_8$ values, the tension with CMB measurements in our extended model space remains at only the $1.75\sigma$ level.

 \item All joint analyses yield $H_0 \sim 67$ km/s/Mpc with sub-percent level uncertainties. Therefore, FS-based joint analyses, even when combined with recent geometric probes, do not alleviate the Hubble tension. This holds true even when allowing for the extended parameter spaces considered in this work.
\end{itemize}

Overall, our results demonstrate that FS data can serve as a competitive and independent cosmological probe, providing strong and consistent support for a late-time phantom crossing scenario. Looking ahead, future public data releases from surveys such as DESI and Euclid \cite{Euclid:2025diy, Euclid:2023tog} are expected to significantly improve the statistical precision of FS measurements. These forthcoming datasets will open new avenues for testing more general theoretical models of dynamical DE and for further exploring potential deviations from the standard cosmological model.

\acknowledgments
\noindent

We thank the referee for their thoughtful comments, which have helped to improve both the clarity and the overall impact of our results. E.S. received support from the CAPES scholarship. R.C.N. thanks the financial support from the Conselho Nacional de Desenvolvimento Científico e Tecnologico (CNPq, National Council for Scientific and Technological Development) under the project No. 304306/2022-3, and the Fundação de Amparo à Pesquisa do Estado do RS (FAPERGS, Research Support Foundation of the State of RS) for partial financial support under the project No. 23/2551-0000848-3.

\appendix
\section{$\Lambda$CDM results}
\label{sec:appendixa}

For comparison purposes, we provide below the table summarizing our results for the standard model. The $S_8$ values systematically remain on the lower side, primarily due to the BOSS DR12 data, which—as previously reported in the literature \cite{Philcox:2021kcw}—tend to favor smaller values.

\begin{table*}[htpb!]
\centering
\Huge
\renewcommand{\arraystretch}{1.8}
\resizebox{\textwidth}{!}{
\begin{tabular}{|l|c|c|c|c|}
\hline
\textbf{Parameter} & \textbf{FS+DESI-DR2} & \textbf{FS+DESI-DR2+PP} & \textbf{FS+DESI-DR2+Union3} & \textbf{FS+DESI-DR2+DESY5} \\ \hline
$10^{2} \Omega_{\rm b} h^2$ & $2.269\pm 0.037$ ($2.269^{+0.073}_{-0.071}$) & $2.270^{+0.035}_{-0.039}$ ($2.270^{+0.075}_{-0.068}$) & $2.268^{+0.036}_{-0.040}$ ($2.268^{+0.078}_{-0.070}$) & $2.271\pm 0.037$ ($2.271^{+0.074}_{-0.070}$)  \\
$\Omega_{\rm c} h^2$ & $0.1184\pm 0.0031$ ($0.1184^{+0.0063}_{-0.0061}$) & $0.1201\pm 0.0031$ ($0.1201^{+0.0063}_{-0.0060}$) & $0.1199^{+0.0029}_{-0.0033}$ ($0.1199^{+0.0063}_{-0.0060}$) & $0.1223\pm 0.0031$ ($0.1223^{+0.0061}_{-0.0061}$)  \\
$n_{s}$ & $0.962\pm 0.041$ ($0.962^{+0.080}_{-0.084}$) & $0.954\pm 0.042$ ($0.954^{+0.080}_{-0.082}$) & $0.953\pm 0.042$ ($0.953^{+0.079}_{-0.085}$) & $0.940\pm 0.041$ ($0.940^{+0.079}_{-0.083}$)  \\
$\ln(10^{10} A_{s})$ & $2.89\pm 0.12$ ($2.89^{+0.23}_{-0.23}$) & $2.87\pm 0.12$ ($2.87^{+0.22}_{-0.22}$) & $2.87\pm 0.12$ ($2.87^{+0.23}_{-0.24}$) & $2.84\pm 0.12$ ($2.84^{+0.22}_{-0.23}$)  \\
$\Omega_{\rm m}$ & $0.2998\pm 0.0056$ ($0.300^{+0.011}_{-0.011}$) & $0.3029\pm 0.0055$ ($0.303^{+0.011}_{-0.011}$) & $0.3027\pm 0.0055$ ($0.303^{+0.011}_{-0.010}$) & $0.3072\pm 0.0054$ ($0.307^{+0.010}_{-0.011}$)  \\
$H_0 \, [\mathrm{km/s/Mpc}]$ & $68.76\pm 0.38$ ($68.76^{+0.74}_{-0.73}$) & $68.81\pm 0.38$ ($68.81^{+0.75}_{-0.71}$) & $68.79\pm 0.38$ ($68.79^{+0.76}_{-0.76}$) & $68.86\pm 0.38$ ($68.86^{+0.75}_{-0.73}$)  \\
$S_8$ & $0.745\pm 0.044$ ($0.745^{+0.090}_{-0.083}$) & $0.748^{+0.042}_{-0.047}$ ($0.748^{+0.090}_{-0.082}$) & $0.746\pm 0.044$ ($0.746^{+0.090}_{-0.083}$) & $0.746\pm 0.044$ ($0.746^{+0.087}_{-0.083}$)  \\
\hline \hline \end{tabular}
}
\caption{$68\%$ and $95\%$ CL constraints for the parameters of the $\Lambda$CDM model.}
\label{table:LCDM}
\end{table*}

\section{Constraining Power of FS in DDE Models}
\label{sec:appendixb}

To ensure greater robustness, we also performed tests excluding the FS data in order to quantify the additional information that FS measurements contribute beyond BAO and SNIa observations. Furthermore, we compared these improvements with those obtained by adding CMB data to the BAO and SNIa baseline, in order to assess whether, for these models, the inclusion of FS data provides gains that are competitive with those from the CMB. As a reference for the CMB-based constraints, we adopted the recent analyses presented in \cite{DESI:2025zgx} and \cite{Scherer:2025esj}.

Our results, summarized in Table~\ref{table:compare}, corroborate the strong constraining power of FS analysis when combined with geometric probes—i.e., without relying on CMB data—consistent also with recent studies \cite{Chen:2025jnr, Ivanov:2024xgb, Noriega:2024eyu, Chudaykin:2020ghx, Philcox:2020vvt, Chudaykin:2025aux}. For the $w_0w_a$CDM model, the first noteworthy point is that, when comparing the results in Table~\ref{table:compare} obtained with FS+DESI-DR2+SNIa to those from \cite{DESI:2025zgx} using CMB+DESI-DR2+SNIa, the error bars on the parameters $w_0$ and $w_a$ are of the same order of magnitude, indicating that the two results are competitive. Regarding $\Omega_{\rm m}$, both the PP and DESY5 combinations yield comparable precision; however, when combined with Union3, the FS-based results are slightly less precise, with FS+DESI-DR2+Union3 giving an uncertainty of 0.012, compared to 0.0086 for CMB+DESI-DR2+Union3.

For the $w_{\dagger}$CDM model, we observe a very similar behavior. When comparing our results with those from \cite{Scherer:2025esj}, we find that the parameter $\Delta$ inferred using CMB+DESI-DR2+SNIa has the same order of magnitude as that obtained in this work with FS+DESI-DR2+SNIa. Similarly, the constraints on $\Omega_{\rm m}$ are also of the same order of magnitude, except for the FS+DESI-DR2+Union3 combination, where CMB+DESI-DR2+Union3 provides tighter precision, with FS-based data yielding $0.321^{+0.010}_{-0.0090}$ compared to $0.321^{+0.0069}_{-0.0077}$ for the CMB-based analysis.

\begin{table*}[htpb!]
\centering
\scriptsize
\renewcommand{\arraystretch}{1.8}
\resizebox{\textwidth}{!}{
\begin{tabular}{|l|c|c|c|c|}
\hline
\textbf{Model/Dataset} & $\mathbf{\Omega_m}$ & $\mathbf{w_0}$ & $\mathbf{w_a}$ & $\mathbf{\Delta}$ \\ \hline
$\mathbf{w_0w_aCDM}$ & & & &\\ 
DESI-DR2+PP & $0.265^{+0.012}_{-0.0096}$ & $-0.846\pm 0.060$ & $-0.56\pm 0.33$ & - \\ 
DESI-DR2+Union3 & $0.338\pm 0.013$ & $-0.64\pm 0.10$ & $-1.33\pm 0.44$ & - \\ 
DESI-DR2+DESY5 & $0.284^{+0.010}_{-0.0091}$ & $-0.675\pm 0.065$ & $-1.24\pm 0.33$ & - \\
FS+DESI-DR2+PP & $0.3181\pm 0.0082$ & $-0.852\pm 0.057$& $-0.60\pm 0.27$ & - \\ 
FS+DESI-DR2+Union3 & $0.338\pm 0.012$ & $-0.670\pm 0.097$ & $-1.23\pm 0.38$ & - \\ 
FS+DESI-DR2+DESY5 & $0.3354\pm 0.0086$ & $-0.692\pm 0.064$ & $-1.18\pm 0.30$ & - \\
$\mathbf{w_{\dag}CDM}$ & & & &\\ 
DESI-DR2+PP & $0.2589^{+0.0084}_{-0.0073}$ & - & - & $-0.084^{+0.030}_{-0.040}$ \\ 
DESI-DR2+Union3 & $0.271\pm 0.010$ & - & - & $-0.151\pm 0.049$\\ 
DESI-DR2+DESY5 & $0.2728\pm 0.0081$ & - & - & $-0.157^{+0.038}_{-0.032}$\\
FS+DESI-DR2+PP & $0.3115\pm 0.0071$ & - & - & $-0.075^{+0.031}_{-0.036}$\\ 
FS+DESI-DR2+Union3 & $0.3217\pm 0.0094$ & - & - & $-0.129^{+0.043}_{-0.048}$\\ 
FS+DESI-DR2+DESY5 & $0.3246\pm 0.0075$ & - & - & $-0.143^{+0.036}_{-0.031}$\\
\hline\end{tabular}
}
\caption{$68\%$ CL constraints for both models considered in this work, with and without the inclusion of FS, on the parameters $\Omega_{\rm m}$, $w_0$, $w_a$, and $\Delta$.}
\label{table:compare}
\end{table*}

\begin{figure*}[htpb!]
    \centering
    \includegraphics[width=0.32\textwidth]{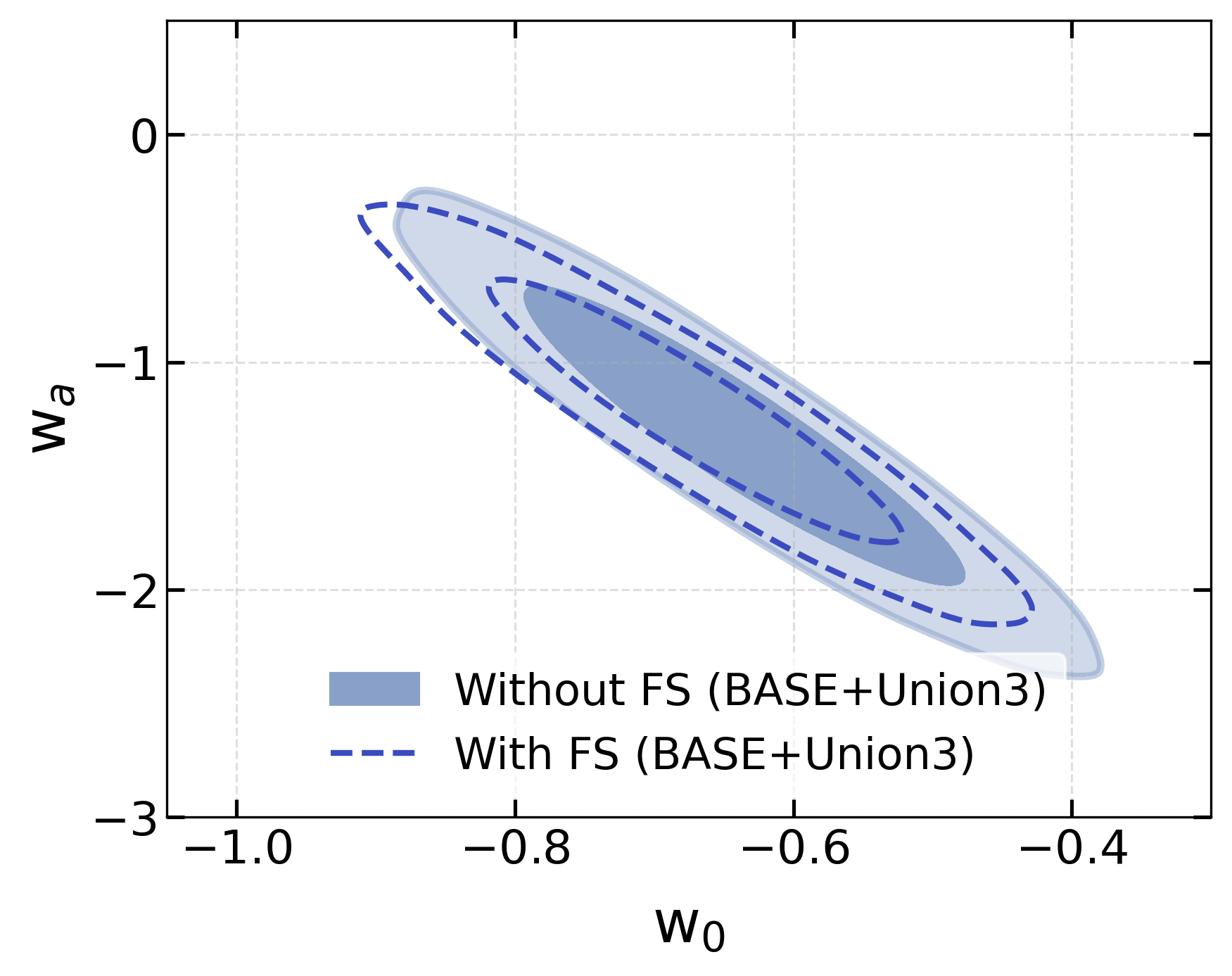}
    \includegraphics[width=0.32\textwidth]{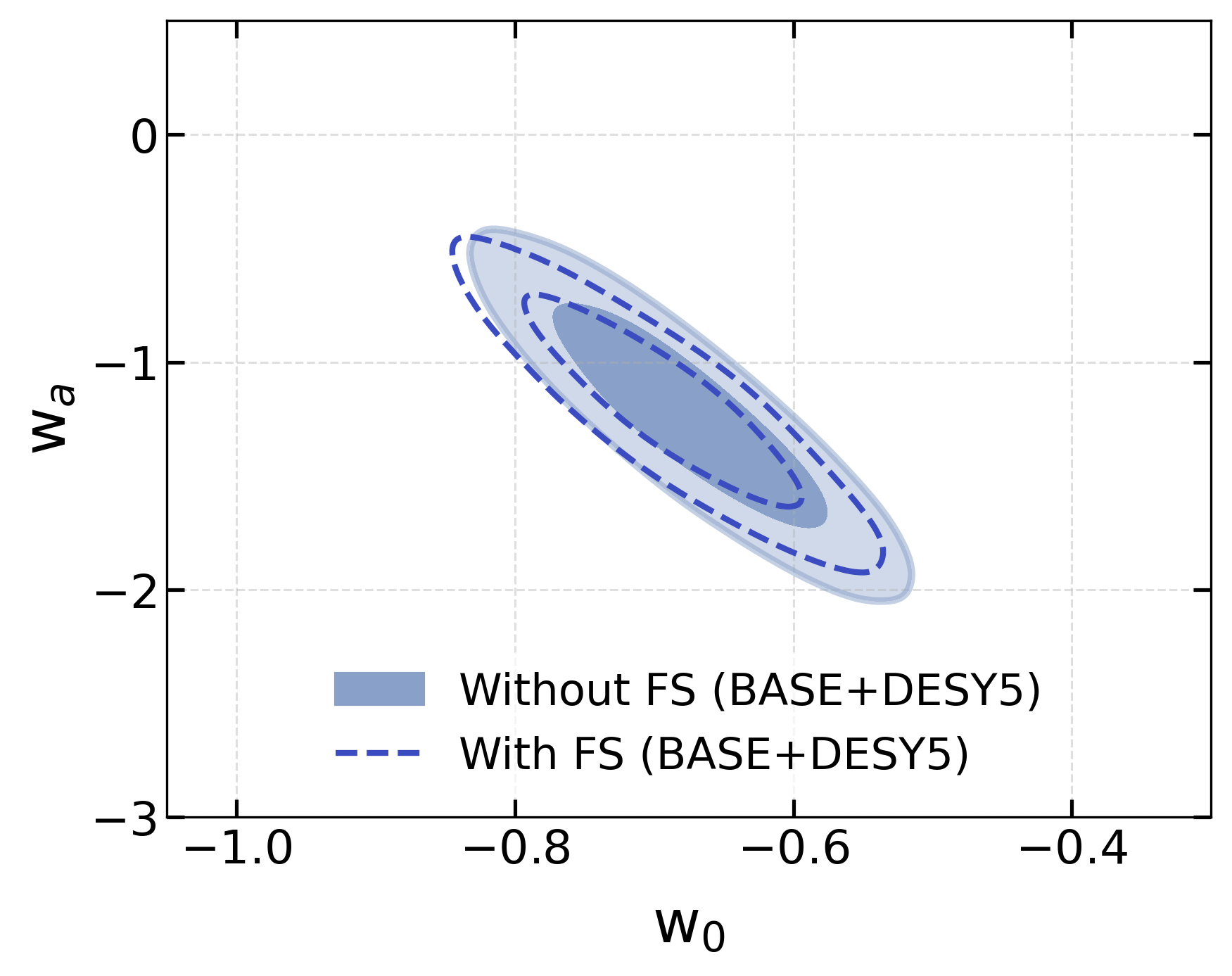}
    \includegraphics[width=0.32\textwidth]{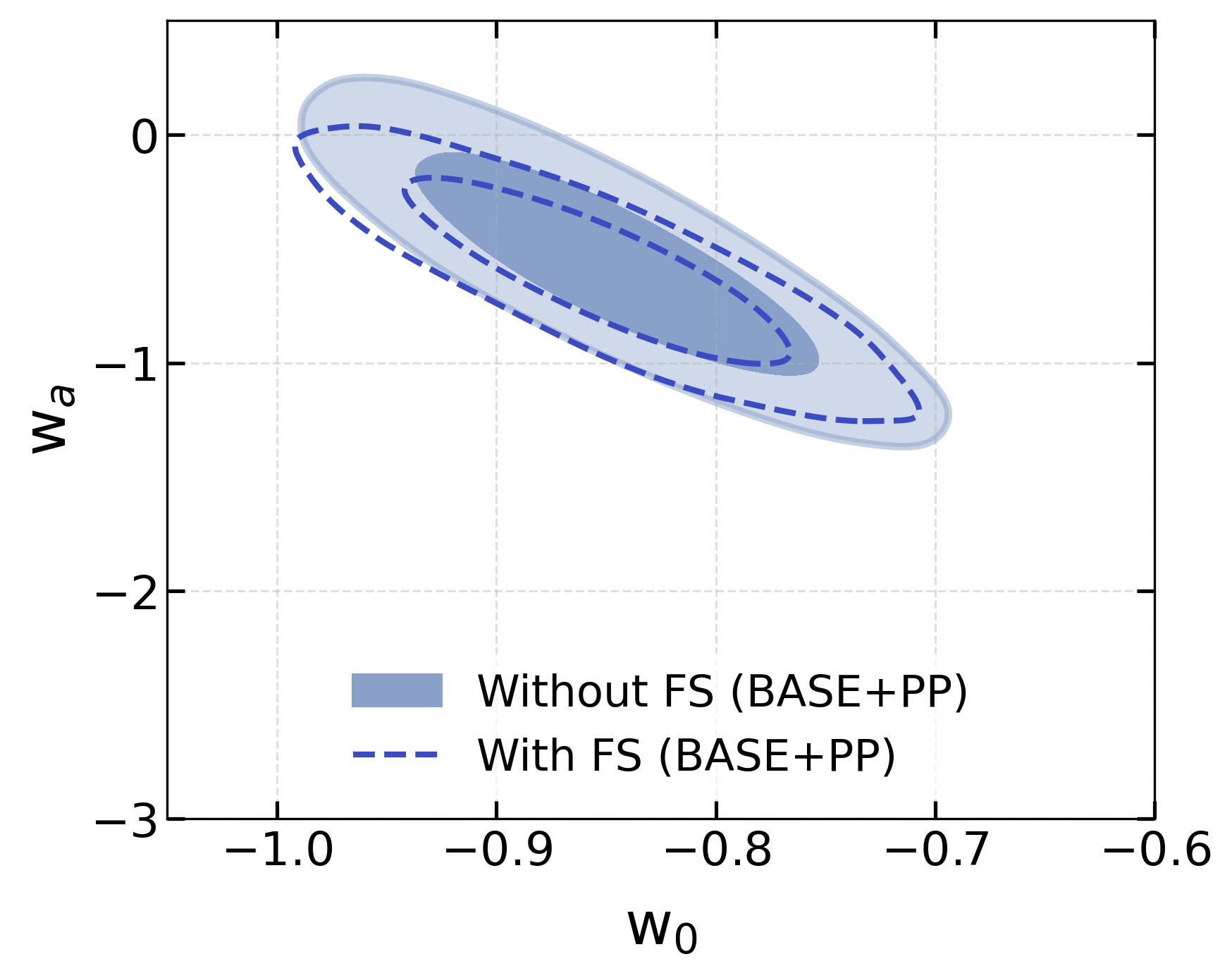}
    \caption{2D posterior distributions for the $w_0w_a$CDM model, shown for each combination of datasets with and without the inclusion of FS data. In the figure legend, the term BASE refers to the DESI-DR2+BBN combination, which is included in all analyses.}
    \label{fig:comparewaw0}
\end{figure*}

Regarding the effective gain in precision, we find that all joint analyses in the $w_0w_a$CDM model benefit from the inclusion of FS information. As shown in Table~\ref{table:compare}, adding the FS likelihood systematically reduces the uncertainties on both $w_0$ and $w_a$. For instance, the DESI-DR2+PP combination shows an improvement of approximately $18\%$ in the constraint on $w_a$ (from $\pm 0.33$ to $\pm 0.27$), accompanied by a comparable refinement in $w_0$ (from $\pm 0.060$ to $\pm 0.057$). A similar trend is observed for DESI-DR2+Union3 and DESI-DR2+DESY5, where the uncertainties on $w_a$ decrease by $\sim 12\%$ and $\sim 9\%$, respectively, once FS information is included. These improvements are consistent with the visible shrinkage of the posterior contours in Fig.~\ref{fig:comparewaw0}, confirming that FS clustering data enhance the constraining power on DDE models.

Beyond DE parameters, the inclusion of FS information also has a significant impact on the determination of the matter density parameter $\Omega_{\rm m}$. As shown in Table~\ref{table:compare}, the addition of FS clustering leads not only to a reduction in the statistical uncertainties but also to a noticeable shift in the central values. For the DESI-DR2+PP combination, the $1\sigma$ uncertainty on $\Omega_{\rm m}$ improves by nearly $30\%$ (from $^{+0.012}_{-0.0096}$ to $\pm 0.0082$), while the mean value shifts from $\Omega{\rm m}=0.265$ to $\Omega_{\rm m}=0.3181$, bringing it into closer agreement with the values obtained from DESI-DR2 combined with the other probes. A similar improvement is observed for DESI-DR2+DESY5, where the uncertainty decreases by $\sim 13\%$ and the central value shifts from $0.284$ to $0.3354$. In contrast, the DESI-DR2+Union3 combination shows a more modest improvement, with the uncertainty on $\Omega_{\rm m}$ decreasing by approximately $8\%$ when FS information is included (from $\pm 0.013$ to $\pm 0.012$). This milder reduction reflects the already strong constraining power of SNIa on the background expansion history, which limits the additional impact of LSS information on $\Omega_{\rm m}$ in this particular combination.

Similar improvements are observed for the $w_{\dagger}$CDM model, where the inclusion of FS information enhances the constraining power of all dataset combinations. As shown in Table~\ref{table:compare}, the uncertainty on $\Omega_{\rm m}$ is most significantly reduced for the DESI-DR2+PP combination, by nearly $15\%$ (from $^{+0.0084}_{-0.0073}$ to $\pm 0.0071$), while more modest reductions of $\sim 6\%$ and $\sim 8\%$ are found for DESI-DR2+Union3 and DESI-DR2+DESY5, respectively. In addition to tightening constraints, FS data induce noticeable shifts in the central values of $\Omega{\rm m}$. These shifts are consistent with those observed in the $w_0w_a$CDM model, reinforcing the robustness of this trend and highlighting the critical role of FS clustering in breaking degeneracies between matter density and DE parameters.

\begin{figure*}[htpb!]
    \centering
    \includegraphics[width=0.32\textwidth]{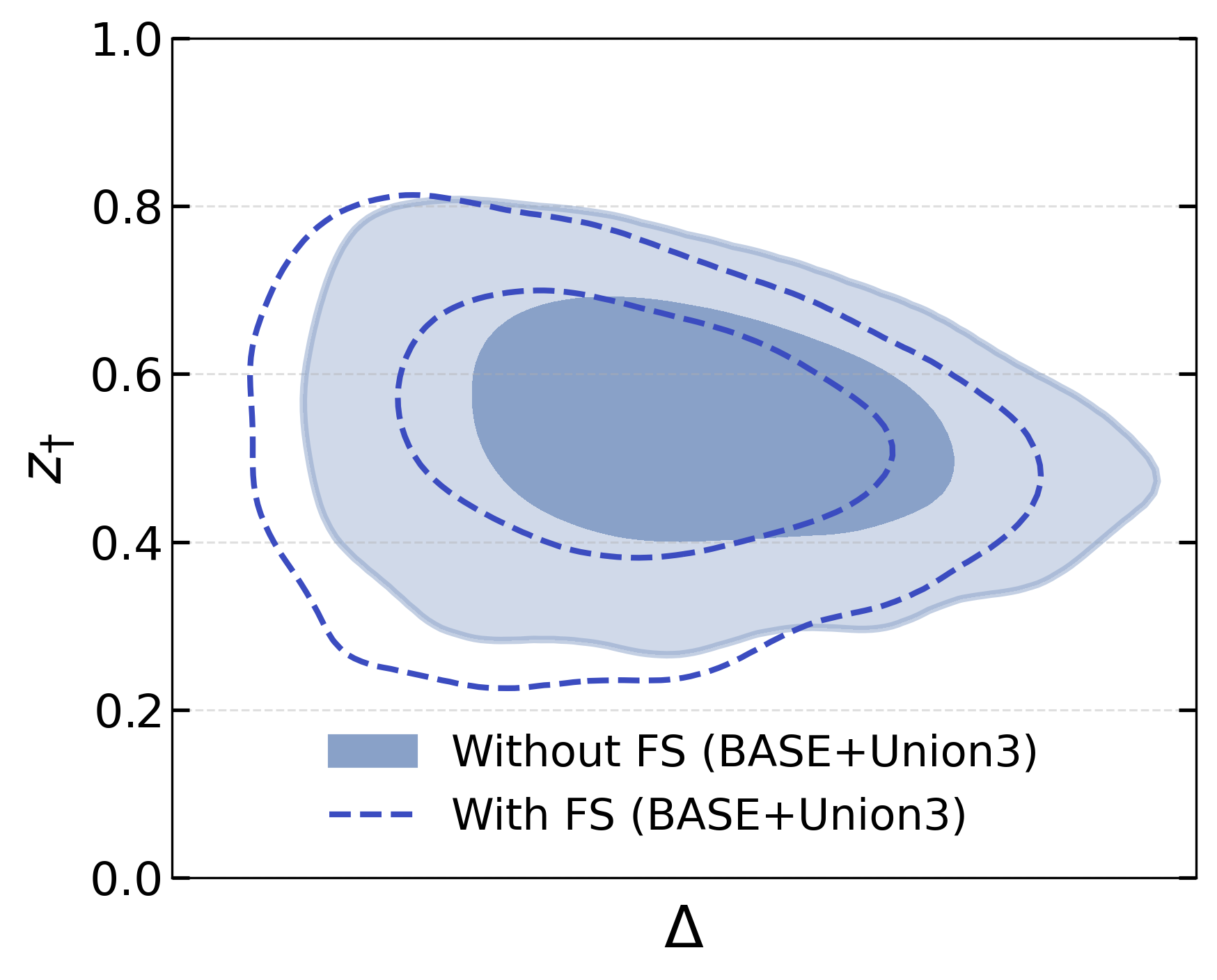}
    \includegraphics[width=0.32\textwidth]{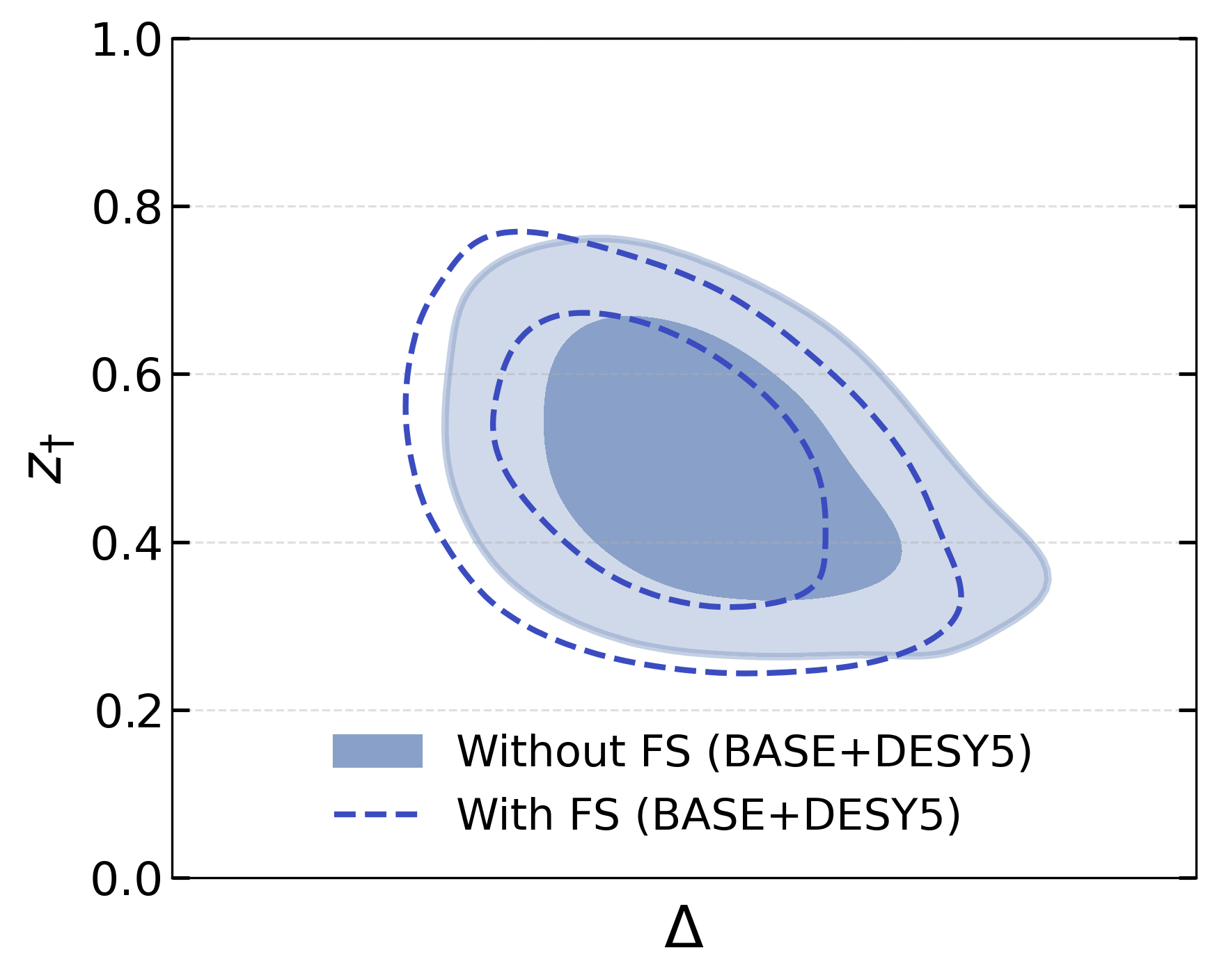}
    \includegraphics[width=0.32\textwidth]{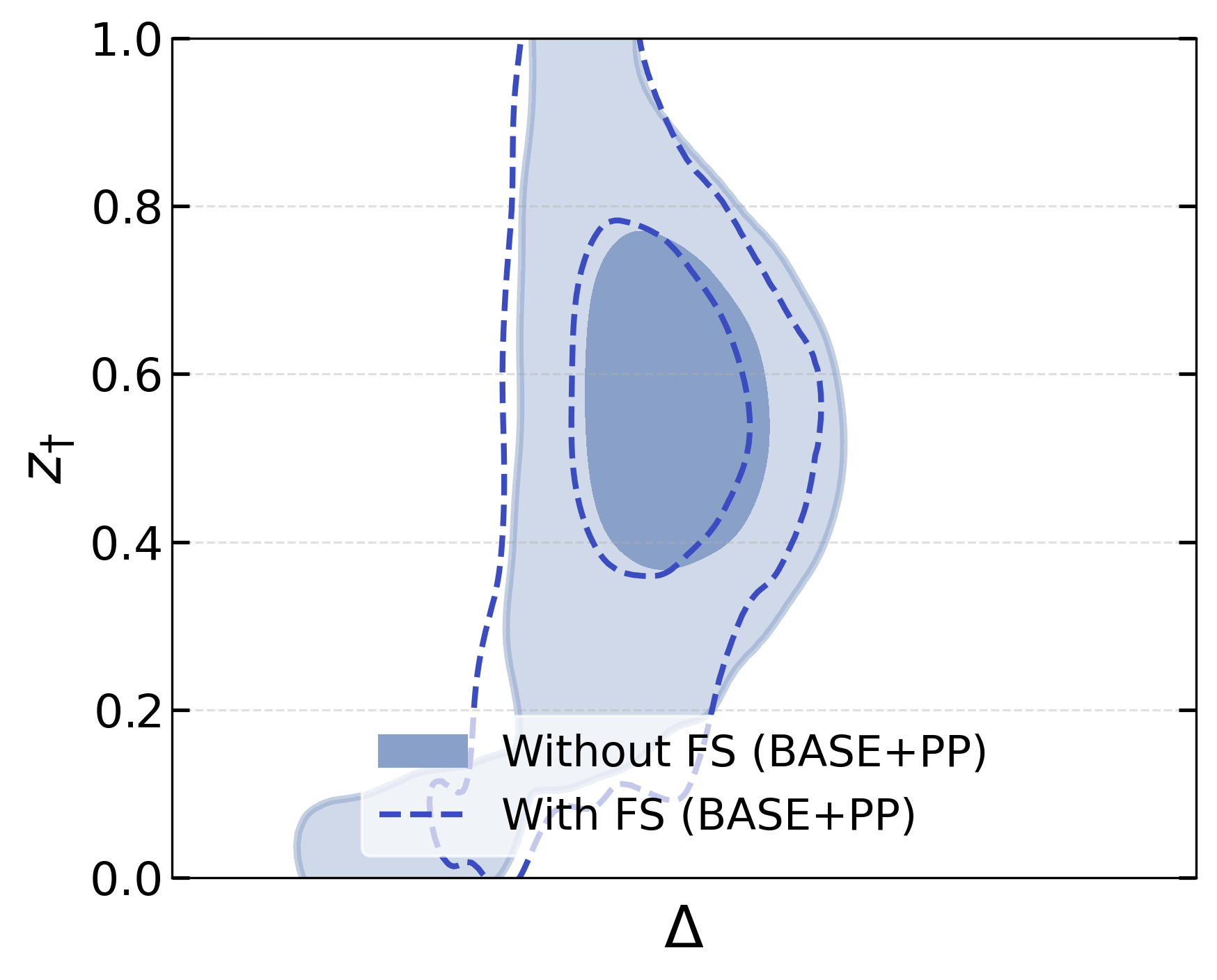}
    \caption{Same as in figure \ref{fig:comparewaw0}, but for the $w_{\dag}$CDM model.}
    \label{fig:comparews}
\end{figure*}

Regarding the parameter $\Delta$, we find moderate improvements, with uncertainties reduced by $\sim 12\%$ for DESI-DR2+Union3, $\sim 8\%$ for DESI-DR2+DESY5, and $\sim 6\%$ for DESI-DR2+PP. These gains are illustrated by the posterior contours in Fig.~\ref{fig:comparews}, where the inclusion of the FS likelihood produces slightly more confined confidence regions.

In general, the inclusion of FS clustering information enhances the constraints on both $w_0w_a$CDM and $w_{\dagger}$CDM models. FS data systematically reduce uncertainties on $w_0$, $w_a$, $\Omega_{\rm m}$, and $\Delta$, while producing consistent shifts in central values that improve agreement across dataset combinations. These improvements are reflected in the posterior contours, demonstrating that FS information provides complementary and robust constraints, breaks degeneracies, and yields a more precise picture of the DE sector.

\bibliographystyle{JHEP}
\bibliography{main}

\end{document}